\documentclass[10pt]{article}  

\usepackage{graphicx}
\usepackage[utf8]{inputenc}
\usepackage[a4paper,left=2cm,right=2.cm, bottom=2.cm, top=2.0cm]{geometry}
\usepackage{amssymb}
\usepackage{amsmath}
\usepackage{bm}
\usepackage[margin=1pt,font=small,labelfont=bf]{caption}
\usepackage[dvipsnames]{xcolor}
\definecolor{citecolor}{RGB}{128,0,32}
\usepackage[unicode, pdfencoding=auto,psdextra,hyperfootnotes=false,breaklinks=true,colorlinks=true,allcolors=citecolor]{hyperref}
\usepackage{xr-hyper}
\usepackage{setspace}
\usepackage{fancyhdr}
\usepackage{titlesec}
\usepackage{algorithm}
\usepackage{algpseudocode}
\usepackage{etoolbox}

\usepackage{times}
\usepackage{tabularx}
\usepackage{datetime} 
\usepackage{lineno}
\usepackage{booktabs}
\usepackage{etoolbox}
\usepackage{subcaption}

\patchcmd{\linenumber}{\hb@xt@}{\hbox}{}{}
\newdateformat{usvardate}{\shortmonthname[\THEMONTH]. \THEDAY, \THEYEAR}

\newcommand{\lastmodified}{%
  \textcolor{gray}{\scriptsize \usvardate\today{} at \currenttime}
}

\fancypagestyle{custom}
{      
\fancyhead[L]{}
\fancyhead[C]{\textcolor{gray}{\scriptsize manuscript submitted to \textit{Journal of Geophysical Research}}}
\fancyhead[R]{}
\fancyfoot[C]{\textcolor{gray}{\thepage}}
\fancyfoot[R]{\lastmodified}
\fancyfoot[L]{}
}
\renewcommand\thesection{\arabic{section}}
\titlespacing\section{0pt}{12pt plus 4pt minus 2pt}{0pt plus 2pt minus 2pt}

\usepackage[
  giveninits=true,
  style=authoryear,
  backend=bibtex,
  maxcitenames=2,
  maxbibnames=99,
  hyperref=true,
  backref=false,
  sorting=nyt,
  uniquename=init,
  autolang=hyphen,
  dashed=false]{biblatex}

\AtEveryCite{\it \color{citecolor}}
\AtEveryBibitem{\clearfield{month}}
\DeclareGraphicsExtensions{.jpg}
\usepackage[breakable]{tcolorbox}
\definecolor{harsha}{RGB}{200,200,255}

\newtcbox{\inlinehl}{
  on line,
  colback=harsha!70,
  colframe=harsha!70,
  boxrule=0pt,
  arc=1mm,
  left=2pt,
  right=2pt,
  top=1pt,
  bottom=1pt,
}

\begin{document}
 
\pagestyle{custom}

\begin{center}
\LARGE {\bf Tidal sensitivity of tremors in a mixed fast and slow earthquake system in northeastern Japan\\[12pt]}
%Tidal sensitivity of tectonic tremor varies with seismicity and tremor migration in northeastern Japan

%Tidal sensitivity of tremor in a multi-perturbation system reveals fault-system state along the Japan Trench
%Statistical analysis on tidal sensitivity of tremors along the Japan Trench associated with clustering, migration and seismicity 
\normalsize

Yishuo Zhou$^{1}$, 
Hideo Aochi$^{1,2}$, 
Alexandre Schubnel$^{1}$, 
Satoshi Ide$^{3}$,
Harsha S. Bhat$^{1}$

\begin{enumerate}
\small
\setlength\itemsep{-5pt}
\item {Laboratoire de G\'{e}ologie, Ecole Normale Sup\'{e}rieure, CNRS-UMR 8538, PSL Research University, Paris, France}
\item {Bureau de Recherches G\'{e}ologiques et Mini\'{e}res (BRGM), 45100 Orl\'{e}ans, France}
\item {Department of Earth and Planetary Science, The University of Tokyo, Tokyo, Japan}
%\item[$\dagger$]Corresponding author: \texttt{yishuozhou1999@gmail.com}
\end{enumerate}

{ \small
\textbf{Keypoints: }
Tidal sensitivity;
shallow tectonic tremor;
Earthquake associated tremor;
Tremor migration;
Japan trench}
\end{center}

\section*{CRediT}

\scriptsize

\begin{tabularx}{\textwidth}{rX}
\textbf{Conceptualization:} & {A. Schubnel, S. Ide, Y. Zhou, H. Aochi, H. S. Bhat}   \\
\textbf{Methodology:} & {Y. Zhou, H. Aochi, A. Schubnel, S. Ide, H. S. Bhat}  \\
\textbf{Software:} & {Y. Zhou} \\
\textbf{Investigation:} & {Y. Zhou} \\
\textbf{Writing -- original draft:} & {Y. Zhou} \\
\textbf{Writing -- review \& editing:} & {H. Aochi, A. Schubnel, S. Ide, H. S. Bhat}  \\
\textbf{Supervision:} & {H. Aochi, A. Schubnel, H. S. Bhat} \\
\textbf{Funding acquisition:} & {A. Schubnel, H. Aochi, H. S. Bhat}
\end{tabularx}

\small

\section*{Abstract}

Tidal modulation of tectonic tremors provides a sensitive measure of fault response to small stress perturbations, yet how this response varies in a mixed fast and slow earthquake system remains unclear. Here we present the first systematic investigation of tremor tidal sensitivity in such a system, focusing on tectonic tremors along the northeastern Japan subduction zone. Using a tremor catalog from 2016 to 2024, we show that the southern end of the Kuril Trench, characterized by tremor migration and relatively weak seismicity, exhibits the strongest tidal sensitivity, whereas the northern Japan Trench shows the weakest response. Spatial analysis further reveals that areas with weaker tidal sensitivity tend to coincide with more earthquakes ($M_j \geq 4$) and denser tremor activity. In addition, tidal sensitivity at the southern end of the Kuril Trench increases from the early to later stages in tremor migration, potentially reflecting changes associated with underlying slow slip processes. Together, these spatial and temporal patterns suggest that tremor tidal sensitivity may be influenced by the relative contribution of other ongoing perturbations. These results highlight tidal sensitivity as a useful probe of the underlying perturbation environment and provide insight into the possible influence of slow slip processes, earthquakes, and other stress changes on tremor-generating regions.

\section*{Plain Language Summary}

Tidal stresses are extremely small, but tectonic tremors are often sensitive to these weak periodic stress perturbation. 
This sensitivity provides information about fault conditions, but how it varies when tremor occurs in a mixed fast and slow earthquake system remains unclear.
Shallow tremors along the northeastern Japan subduction zone provide an ideal opportunity to investigate this question. 
Although these tremors occur outside the main coseismic slip region of the 2011 Mw 9.0 Tohoku-oki earthquake, some tremor regions spatially overlap with other earthquakes, including events larger than magnitude 6. Clear tremor migration, which may be associated with slow slip processes, is also observed in some regions. We find that areas with weaker tidal sensitivity are spatial coincidence with areas with more earthquakes and more tremors. And tidal sensitivity higher in the later stages of tremor migration, possibly reflecting changes associated with slow slip processes. 
Together, these observations suggest that tidal sensitivity may provide a window into the underlying perturbation environment of tremor generating regions. It may help reveal the potential influence of earthquakes, slow slip processes, and other stress perturbations on tremor generating regions along subduction zones.

\section*{Key Points}
\begin{itemize}

\item Tidal sensitivity of tremors is systematically investigated for the first time in a mixed fast and slow earthquake system.
\item Areas with weaker tidal sensitivity spatially coincide with more earthquakes ($M_j \geq 4$) and denser tremor activity.
\item Tidal sensitivity may provide a useful probe for perturbation environments in tremor generating regions.

\end{itemize}

\normalsize
\section{Introduction}
Slow earthquakes have revealed that fault slip along subduction interfaces can occur in a much broader range of styles than the simple contrast between stable creep and regular earthquakes. These phenomena, including slow slip events (SSEs), tectonic tremor, low-frequency earthquakes (LFEs) and very low-frequency earthquakes (VLFEs), provide important constraints on fault slip processes and on the physical conditions governing megathrust behavior \parencite{kato2012propagation, ito2013episodic,obara2025slow,shelly2026low}. Among them, tectonic tremor is particularly useful because it is detectable in seismic data and provides abundant observations of slow fault slip activity. Tremor often occurs in migrating bursts along the plate interface and is commonly observed during short-term slow slip (ETS), making it a useful window into the evolution of fault slip \parencite{rogers2003episodic}.

Tectonic tremor is widely observed to be sensitive to tidal stresses despite their extremely small amplitudes ($\sim$kPa), and this sensitivity has been used to constrain the stress conditions in the tremor source region, particularly the effective normal stress \parencite{shelly2007complex,rubinstein2008tidal,nakata2008non,thomas2009tremor,lambert2009correlation,ide2015thrust,houston2015low,chen2018tidal,yi2025characteristics}. Some studies have reported that tidal sensitivity increases during the later stages of migrating tremor clusters in the Nankai and Cascadia subduction zones \parencite{thomas2013evidence,yabe2015tidal,houston2015low,peng2015high,katakami2017tidal}. Enhanced tidal sensitivity has also been documented during long-term slow slip events in the Nankai subduction zone \parencite{hirose2025tidal}. These observations suggest that tidal sensitivity may reflect evolving fault conditions associated with slow slip processes. However, the potential influence of other coexisting stress perturbations, especially nearby fast earthquake, on tidal sensitivity of tremors has not been investigated.

\begin{figure}
    \centering
\includegraphics[width=0.5\linewidth]{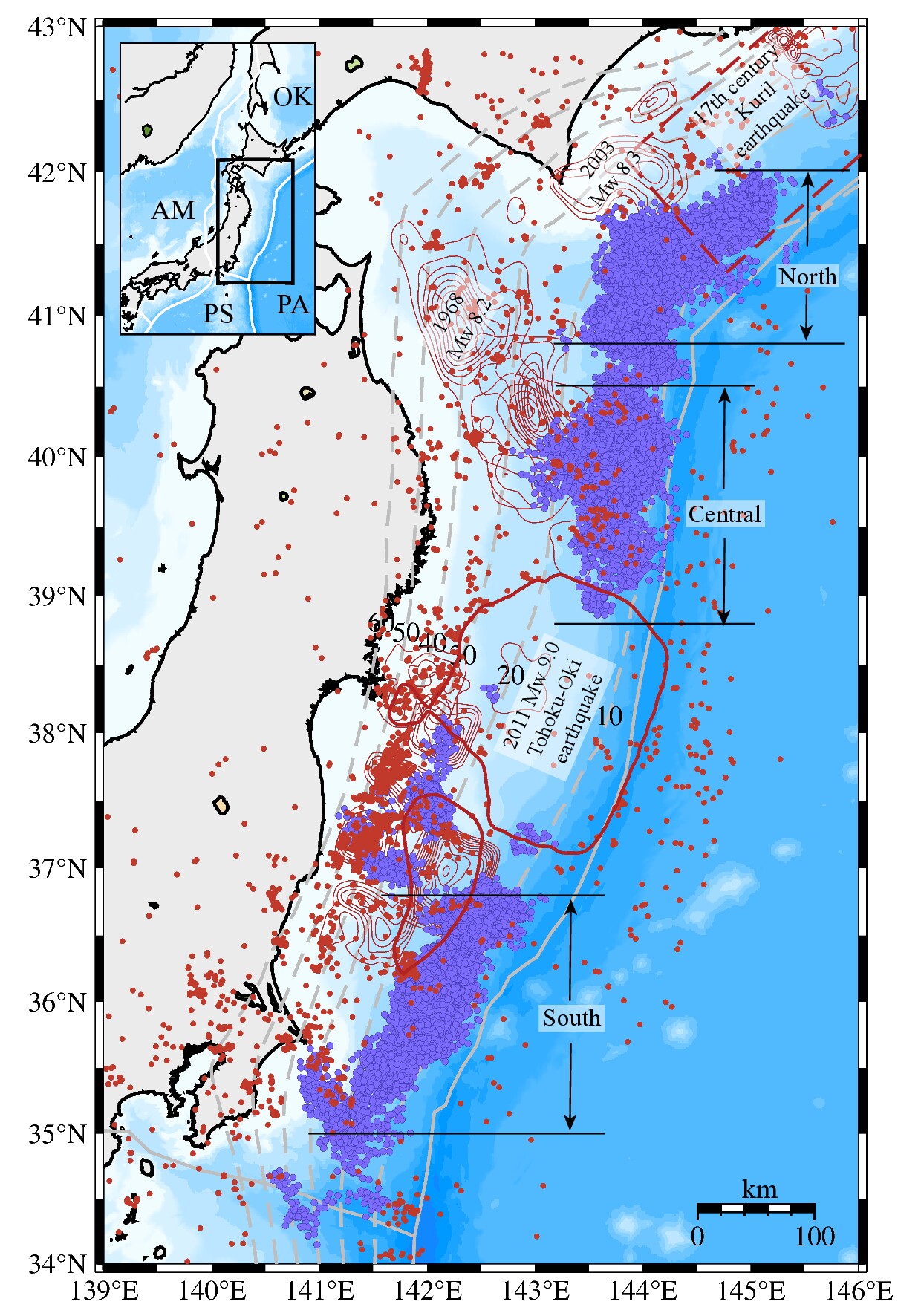}
    \caption{Spatial distribution of tectonic tremors (purple dots) along the Japan Trench between 14 August 2016 and 31 August 2024 \parencite{sagae2025machine}. Fast earthquakes (red dots; only earthquakes with $M_j \geq 4$ are shown for clarity) from the Japan Meteorological Agency (JMA) catalog. Red contour lines denote published outlines of coseismic high-slip region for the 2011 $M_\mathrm{w}$~9.0 Tohoku-oki earthquake (10 m) \parencite{iinuma2012coseismic} and other major megathrust earthquakes along the trench, compiled from previous source models \parencite{utokyo_eic_2003_tokachi_oki,utokyo_eic_2004_off_kushiro,utokyo_eic_2005_miyagi_oki,yamanaka2004asperity,murotani2003rupture,nagai2001comparative,yamanaka2003source}. The red dashed rectangle represents the fault model of the 17th century Kuril earthquake \parencite{ioki2016re}. The study region is subdivided into three along-strike regions (North, Central, and South). The light gray line indicates the plate boundary \parencite{bird2003updated}. Gray dashed lines denote depth contours (10–60 km) of the subducting Pacific plate interface \parencite{baba20063}. The rectangle in the inset with the plate boundary \parencite{bird2003updated} represents the location of the study area. PA Pacifc plate, PS Philippine Sea Plate, AM Amur plate, OK Okhotsk plate.
}
    \label{fig:study_map}
\end{figure}

The northeastern Japan subduction zone (34--43$^\circ$N) provides an ideal natural setting to study this question (Figure~\ref{fig:study_map}). Along the Kuril-Japan trench system, where the Pacific Plate subducts beneath the Okhotsk plate, Mw 7-8 interplate megathrust earthquakes have occurred at depths of 20-50 km between 37$^\circ$N and 43$^\circ$N along the trench \parencite{nagai2001comparative,murotani2003rupture,yamanaka2004asperity,yagi2004source}. The 2011 Mw 9.0 Tohoku-Oki earthquake off 37--39$^\circ$N ruptured to the trench axis with more than 50~m of slip \parencite{ide2011shallow,simons20112011}, and 17th century earthquake (Mw $\sim$9) off the southern Kuril subduction zone are also inferred to maybe have involved substantial shallow megathrust slip \parencite{nanayama2003unusually}. Slow earthquake activity along the Japan Trench was first reported in 1992 as the Sanriku-Oki ultraslow earthquake, a large transient aseismic slip event at depths of approximately 10--20 km (39--40$^\circ$N) \parencite{kawasaki19951992}. Subsequent studies further identified other types of shallow slow earthquakes in this region.  Tectonic tremors and VLFEs have since been identified along the shallow plate interface at depths of 10--20 km \parencite{ohta2019tremor,tanaka2019shallow,nishikawa2019slow,baba2020comprehensive}. Their distribution is interrupted in 37--39$^\circ$N, forming a tremor gap that corresponds to the large coseismic slip area of the 2011 Tohoku-Oki earthquake \parencite{iinuma2012coseismic,nishikawa2019slow}. More recently, deep-learning-based detection has revealed additional shallow tectonic tremors, further expanding the known tremor distribution in this region\parencite{sagae2025machine}. Along-strike tremor migration with a velocity of $\sim$10 km/day has been observed in the north of $\sim$41$^\circ$N \parencite{tanaka2019shallow,nishikawa2019slow,nishikawa2023review}. Although short-term SSEs have been reported at depths of 10-30 km and 40-60 km \parencite{nishimura2021slow}, their catalog may remain incomplete near the trench axis, whereas tectonic tremors are only clearly observed on the shallow plate interface. 
Unlike many other subduction zones where slow earthquakes and fast earthquakes are more clearly separated, the Japan Trench represents a mixed fast and slow earthquake system, where shallow tectonic tremor and fast earthquake activity are closely distributed, and locally overlap, in map view \parencite{nishikawa2023review}. Recent work by \textcite{farge2025big} suggested that even nearby small earthquakes can perturb the spatial synchronization of tectonic tremor for the global catalog including northeastern japan, indicating that earthquake-related perturbations may influence tremor organization. For tidal sensitivity, \textcite{tanaka2012tidal} reported the tidal modulation of Mw $\geq$ 5 earthquakes in 37--39$^\circ$N before the 2011 Tohoku-Oki earthquake, suggesting a critically stressed source region. However, the tidal sensitivity of shallow tectonic tremor in such a mixed fast and slow earthquake system has not yet been investigated. The Japan Trench therefore provides an ideal setting to examine whether tremor tidal sensitivity is also influenced by nearby fast earthquake activity.

Here, we present the first systematic study of tremor tidal sensitivity in a mixed fast and slow earthquake system, using a recently published tremor catalog along the northeastern Japan subduction zone from 2016 to 2024 \parencite{sagae2025machine}.
We first characterize along-strike variations in tremor and earthquake activity. 
We then quantify tidal sensitivity along the Japan Trench and examine its spatial relationship with earthquakes and tremors numbers. 
Next, we investigate regional behavior in more detail, including earthquake-associated tremor rate changes, tidal sensitivity changes in selected regions, and tidal sensitivity changes during tremor migration.
Finally, we discuss possible factors that may contribute to these regional differences.

\section{Data and analysis method}

\subsection{Tremor catalog}
The tremor catalog developed by \textcite{sagae2025machine} contains tectonic tremor detections along the offshore northeastern Japan subduction zone from 15th August 2016 to 31th August 2024. It was constructed from continuous waveform data recorded by the dense S-net cable-type ocean-bottom seismometer network \parencite{aoi2020mowlas}, using a machine-learning-based tremor monitoring system that classifies 1-min waveform windows and associates multi-station detections to determine tremor locations. According to \textcite{sagae2025machine}, the horizontal location uncertainties of most tremor events are within \(\sim\)10~km, whereas depth uncertainties are generally larger. We therefore use only the tremor origin times and epicentral locations from the catalog, and assume that tremors occur on the plate interface for tidal stress calculations. Guided by the broad along-strike distribution of tectonic tremor reported in previous studies \parencite[e.g.,][]{nishikawa2023review,sagae2025machine}, we focus on three major tremor regions in the northeastern Japan subduction zone. We define the southern end of the Kuril Trench (40.8--42$^\circ$N; hereafter the Northern region), the northern Japan Trench (38.8--40.5$^\circ$N; Central region), and the southern Japan Trench (35.0--36.8$^\circ$N; Southern region). These three regions contain 125{,}830 tremor events during the study period.

\subsection{Fast earthquake catalog}
Earthquake data are obtained from the Japan Meteorological Agency (JMA) unified catalog through the end of 2023. Events after 2023 are supplemented using the JMA daily earthquake reports, forming a continuous catalog for the study period. We compile earthquakes with magnitudes $M_j \geq 2$ within the three study regions, yielding a total of 18{,}752 events over the study period. The specific magnitude threshold, spatial window, and temporal window used in each analysis are described in the relevant sections.

\subsection{Tidal stress calculation}

\begin{figure}
    \centering
    \includegraphics[width=0.5\linewidth]{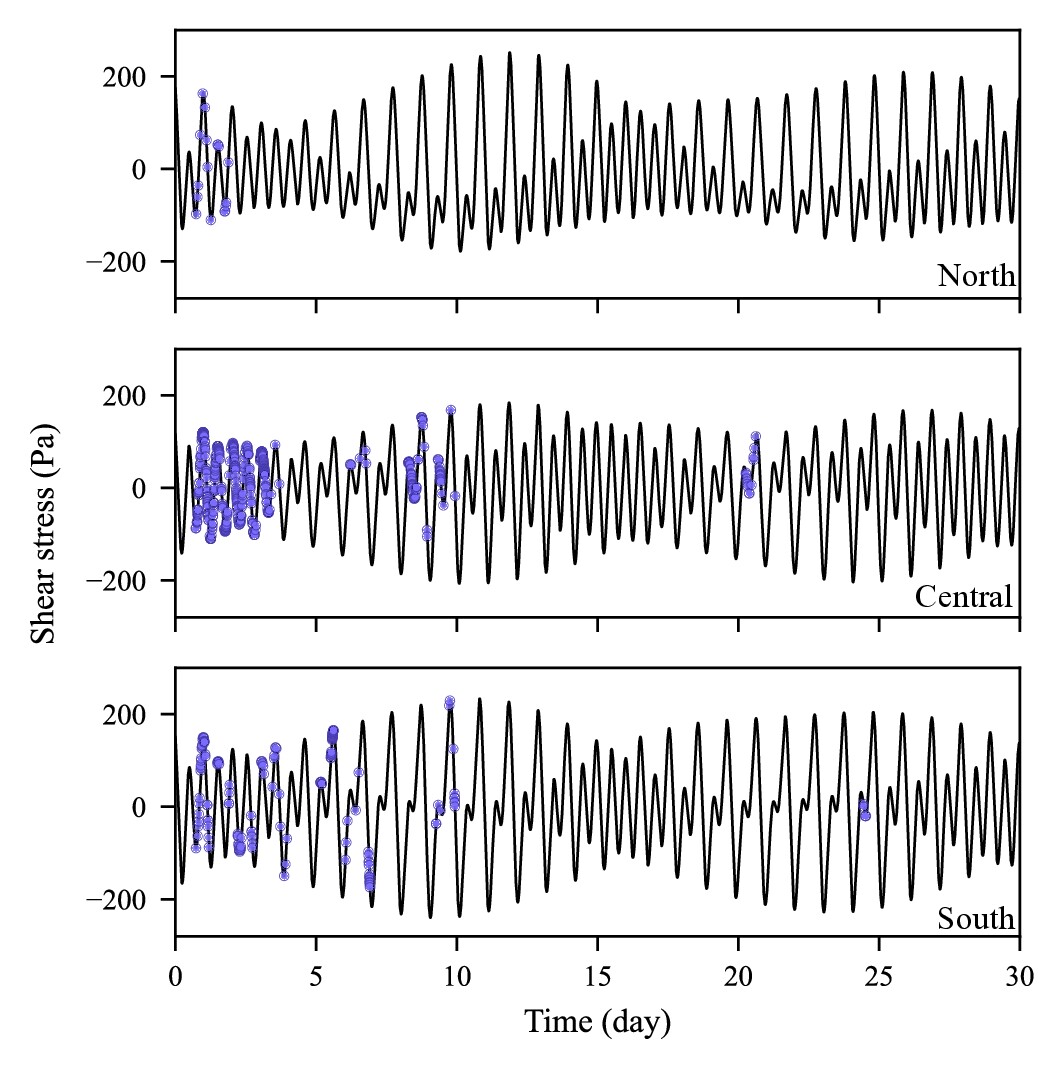}
    \caption{
Examples of calculated tidal shear stress at the centers of the northern, central, and Southern regions. Purple dots indicate tremor occurrence times.}
    \label{fig:tidal_shear_stress}
\end{figure}

Tidal stresses are calculated for the three study regions using the implementation of \textcite{yabe2015tidal}, which considers both body tides and ocean tidal loading. For each region, tidal stress is evaluated at the regional center, defined as the mean latitude and longitude of tremor locations. In this implementation, body tides are computed from the tidal potential of \textcite{tamura1987harmonic}. Ocean tidal loading is calculated using SPOTL \parencite{agnew2012spotl} with the global model NAO99b and the regional model NAO99Jb around Japan \parencite{matsumoto2000ocean}. Ten tidal constituents (M2, S2, N2, K2, K1, O1, P1, Q1, MF, and MM) are included. Green's functions are computed following \textcite{okubo2001complex} using PREM \parencite{dziewonski1981preliminary}. To resolve tidal stresses onto the fault plane of tremor sources, the fault orientation (strike, dip, and rake) need to be specified. Because focal mechanisms of tremor are not directly constrained, we adopt representative fault orientations based on slab geometry. Specifically, the dip direction is taken as the direction of maximum slab gradient from the slab geometry model of \textcite{baba20063}, and the rake is determined by projecting the relative Pacific--Eurasia plate motion from MORVEL onto the fault plane. For the Northern, Central, and Southern regions, we use representative focal mechanisms of (strike, dip, rake) = (221$^\circ$, 6$^\circ$, 70$^\circ$), (188$^\circ$, 5$^\circ$, 101$^\circ$), and (207$^\circ$, 9$^\circ$, 82$^\circ$), evaluated at regional centers of (41.36$^\circ$N, 144.17$^\circ$E), (40.01$^\circ$N, 143.67$^\circ$E), and (36.03$^\circ$N, 142.07$^\circ$E), respectively. Tidal stresses are calculated at 6-min intervals, and the tidal stress and phase at tremor origin times are obtained by linear interpolation. Figure~\ref{fig:tidal_shear_stress} shows examples of the calculated tidal shear stress at the centers of the three regions. Temporal variations are similar among regions, with peak amplitudes of approximately $200$~Pa. The corresponding tidal normal stresses are of order 10~kPa (Figure S1 in Supporting Information).

\subsection{Tidal sensitivity analysis}

In the main text, tidal sensitivity is evaluated with respect to tidal shear stress. Corresponding analyses based on tidal normal stress are presented in the Supplementary Material and show consistent patterns. Following the framework of \textcite{lu2025exploring}, the tidal phase and stress amplitude at the origin time of each tremor are calculated, and tidal sensitivity is quantified using two complementary methods: (1) phase-based analysis \parencite{xue2025probing,zhao2025tidal} and (2) amplitude-based analysis \parencite{houston2015low,yabe2015tidal,yi2025characteristics,hirose2025tidal}.

\subsubsection{Phase-based analysis}

The phase-based analysis evaluates whether tremor occurrence is modulated by tidal phase. For each tremor, the tidal phase is defined relative to the nearest stress maximum, which is assigned $0^\circ$, while adjacent minima correspond to $\pm180^\circ$. The phase between successive extrema is linearly interpolated in time. After assigning a phase to each tremor event, the phase range from $-180^\circ$ to $180^\circ$ is divided into 19 phase intervals: the two edge intervals near $\pm180^\circ$ have widths of $10^\circ$, whereas the 17 interior intervals have widths of $20^\circ$. The resulting phase dependence is fitted with

\begin{equation}
\frac{P_{\mathrm{obs}}(\phi)}{P_{\mathrm{ref}}(\phi)}
= 1+ P \cos(\phi - \phi_0),
\end{equation}

where $P_{\mathrm{obs}}(\phi)$ is the observed tremor occurrence frequency at phase $\phi$, and $P_{\mathrm{ref}}(\phi)$ is the corresponding reference frequency expected under uniform temporal occurrence. The parameter $P$ is the modulation amplitude, with larger values indicating stronger tidal modulation and $P=0$ corresponding to no modulation, that is, temporally uniform tremor occurrence. The parameter $\phi_0$ denotes the phase shift of the fitted peak relative to the tidal stress maximum; $\phi_0 = 0^\circ$ indicates peak tremor occurrence at the tidal stress maximum, whereas $\phi_0 = -90^\circ$ corresponds to peak occurrence near the maximum stressing rate. The parameters $P$ and $\phi_0$ are estimated by least-squares fitting to the phase distribution.

\subsubsection{Amplitude-based analysis}

The amplitude-based analysis evaluates whether tremor occurrence is modulated by tidal stress amplitude. For each tremor event, the tidal shear stress at the origin time is calculated. The stress range is divided into 19 equally spaced bins. The resulting stress dependence is fitted with

\begin{equation}
\frac{P_{\mathrm{obs}}(\tau)}
     {P_{\mathrm{ref}}(\tau)}
= C \exp(\alpha \tau),
\label{Eq:tidal_amplitude}
\end{equation}

where $P_{\mathrm{obs}}(\tau)$ and $P_{\mathrm{ref}}(\tau)$ denote the normalized frequencies in each stress bin at stress $\tau$ for observed tremor times and reference times, respectively. Specifically, the frequency is defined as the number of samples in a given stress bin divided by the total number of samples. Here, $P_{\mathrm{ref}}(\tau)$ represents the expected stress distribution for uniformly occurring events in time. Because the tidal stress time series is irregular and does not spend equal time at all stress levels, randomly timed events would already produce a non-uniform stress distribution. The ratio $P_{\mathrm{obs}}(\tau)/P_{\mathrm{ref}}(\tau)$ therefore corrects for this background sampling bias and quantifies the amplitude-dependent modulation of tremor occurrence.
 The variable $\tau$ denotes the tidal shear stress (in kPa). The parameter $\alpha$ (kPa$^{-1}$) quantifies tidal sensitivity, with $\alpha = 0$ indicating no dependence on stress, and larger values of $\alpha>0$ corresponding to stronger sensitivity. Negative values of $\alpha$ would imply preferential occurrence under stress conditions opposing failure and are generally not physically meaningful. The parameter $C$ is a normalization constant.

\section{Spatiotemporal distribution of tremors and earthquakes}

\begin{figure}
    \centering
    \includegraphics[width=0.9\linewidth]{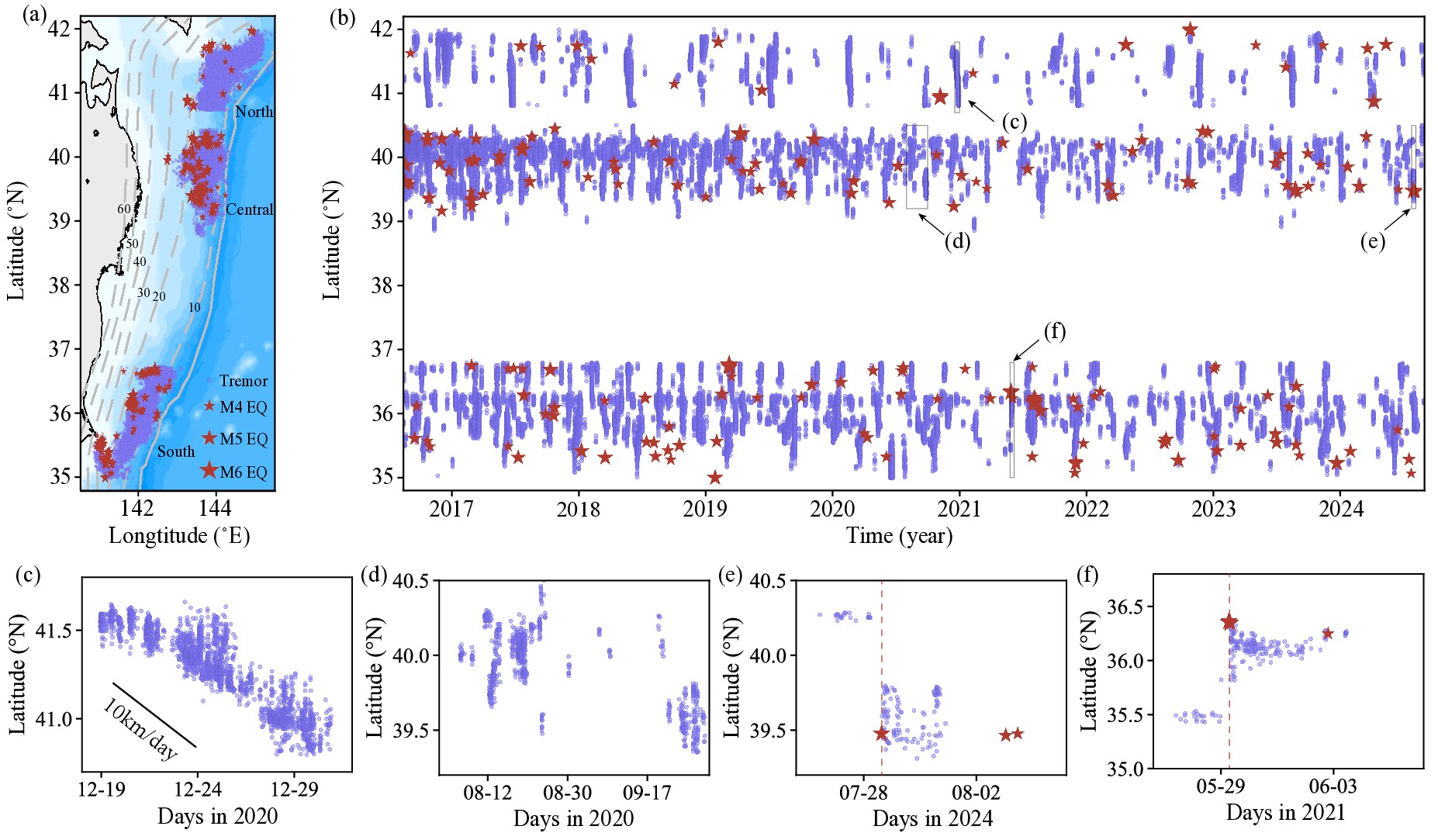}
    \caption{Spatiotemporal distribution of tremors and earthquakes from August 2016 to August 2024. (a) Map view of tremor locations (purple dots) and fast earthquakes with magnitudes $M_j \geq 4$ (red stars). Earthquake data are from the Japan Meteorological Agency (JMA) catalog. The upper surface of the subducting Pacific Plate is shown by gray contours at 10-km depth intervals. (b) Time-latitude distribution of tremor and earthquake occurrences. (c)-(h) Enlarged views of selected portions of panel (b), shown to better illustrate the local spatiotemporal evolution of tremor and earthquake activity. For visual clarity, only earthquakes with $M_j \geq 4$ are shown. To emphasize the spatial association between tremor and seismicity, only earthquakes located in $0.1^\circ \times 0.1^\circ$ grid cells containing tremor are included.}
    \label{fig:spatiotemporal}
\end{figure}

We first examine the spatiotemporal distribution of tremor and nearby earthquakes in the three along-strike regions to establish the baseline characteristics of seismic activity. Figure~\ref{fig:spatiotemporal} shows the temporal evolution and spatial distribution of tremors and earthquakes. Earthquake rate densities reported in this section are calculated using the $M_j \geq 2$ catalog.

In the Northern region, consistent with the migrating tremor behavior reported in previous studies \parencite{nishikawa2023review}, tremor commonly occurs as spatially and temporally coherent clusters, typically lasting on the order of 10-20 days and recurring roughly annually. Our analysis, extended through 2024, indicates that this migration style remains a persistent characteristic of the Northern region. As shown by the representative example in Figure~\ref{fig:spatiotemporal}c, one cluster migrates predominantly southward in latitude, from about $41.6^\circ$N to $40.9^\circ$N over approximately 12 days, corresponding to a migration velocity of about 10~km/day. Seismicity is relatively weak in this region, with an earthquake rate density of about 40~yr$^{-1}$~deg$^{-2}$.

In the Central region, the number of tremors gradually decreases, and tremors tend to occur in multiple spatially separated clusters rather than forming a single coherent migrating front (Figure~\ref{fig:spatiotemporal}d). Seismicity is stronger than in the Northern region, with an earthquake rate density of about 170~yr$^{-1}$~deg$^{-2}$. Many earthquakes spatially overlap with tremor in map view (Figure~\ref{fig:spatiotemporal}a), and we observe the cases in which tremor suddenly increases following nearby earthquakes (e.g., Figure~\ref{fig:spatiotemporal}e). A detailed analysis of changes in tremor rate before and after earthquakes and their spatial relationship will be presented in Section~\ref{sec_eq_tr}.

In the Southern region, tremor is also dispersed along latitude (shown in Figure~\ref{fig:spatiotemporal}b). This region has the highest earthquake rate density, about 260~yr$^{-1}$~deg$^{-2}$. Earthquakes tend to recur at localized sites near the edges of the tremor region, and and cases of sudden tremor activation following nearby earthquakes are also observed (e.g., Figure~\ref{fig:spatiotemporal}f).

Overall, the three regions exhibit distinct tremor and earthquake characteristics. 
The Northern region is characterized by tremor migration with relatively weak seismicity, whereas the Central region shows stronger spatial coexistence between tremors and earthquakes. 
In the Southern region, earthquakes are mainly concentrated near the edges of the tremor zone. 
These contrasting behaviors indicate different perturbation environments along-strike, motivating a comparison of their tidal responses in the following section.

\section{Spatial variations in tremors tidal sensitivity}
\subsection{Along-strike variations}

\begin{figure}[!h]
    \centering
    \includegraphics[width=\textwidth]{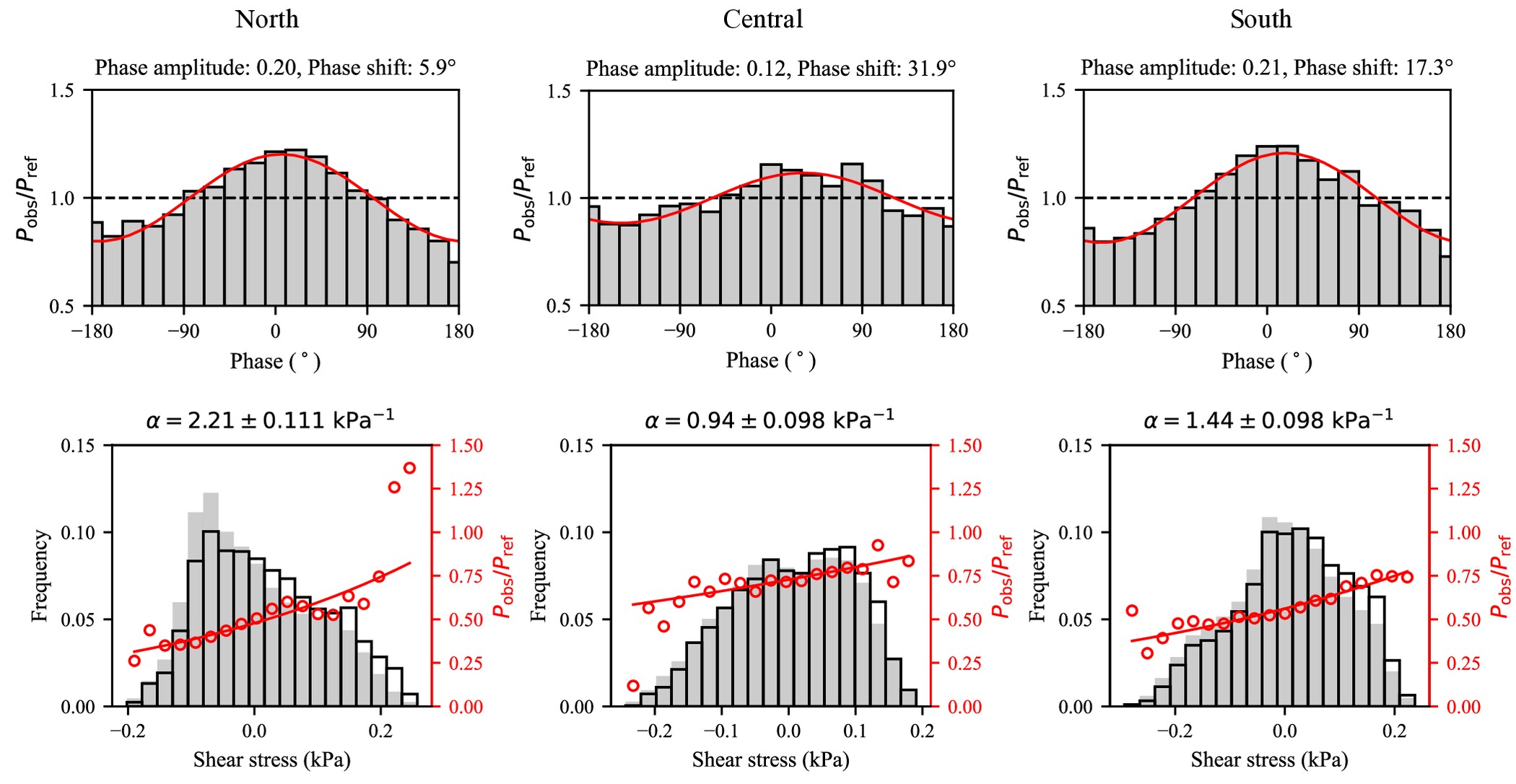}
    \caption{Tidal sensitivity of tremor occurrence in the northern, central, and Southern regions based on tidal shear stress. Tidal shear phase and stress amplitude are assigned to both observed tremor events ($N_{\mathrm{obs}}$) and reference events ($N_{\mathrm{ref}}$). Reference events are sampled at 1-h intervals to represent the background temporal distribution and are used to normalize the observed tremor occurrence. Upper panels show the phase dependence of tremor occurrence, with phase binned every $20^\circ$ (19 bins from $-180^\circ$ to $180^\circ$). The vertical axis shows the ratio of observed to reference event counts in each phase bin. The red curve is a cosine fit to the normalized phase distribution; its amplitude $P$ measures the strength of tidal modulation, and its phase shift indicates the offset of the tremor occurrence peak from the tidal shear stress maximum ($0^\circ$). Lower panels show the dependence of tremor occurrence on tidal shear stress amplitude, with stress divided into 19 equally spaced bins. Gray and red histograms denote the stress distributions of reference and observed events, respectively, and black circles show the normalized observed-to-reference rate. The red curve is an exponential fit to the normalized rate, and the parameter $\alpha$ quantifies the sensitivity of tremor occurrence to tidal shear stress. Corresponding results for tidal normal stress are shown in the Supplementary Material and display the same regional pattern.}
    \label{fig:three region tidal sen}
\end{figure}

We first examine the overall tidal sensitivity of tremor occurrence in the Northern, Central, and Southern regions. Figure~\ref{fig:three region tidal sen} shows the tidal modulation of tremor occurrence in the three regions, quantified by two complementary measures of tidal sensitivity: the phase-based modulation amplitude $P$ and the stress-based sensitivity coefficient $\alpha$.

In the phase-based analysis, the modulation amplitude $P$ is 0.20 in the Northern region, 0.12 in the Central region, and 0.21 in the Southern region, indicating weaker periodic modulation in the Central region than in the other regions. The corresponding phase shifts are $5.9^\circ$, $31.9^\circ$, and $17.3^\circ$, respectively, showing that tremor occurrence in all three regions is most likely near the tidal stress peak, although the preferred phase differs slightly among regions.

The amplitude-based analysis yields a consistent result. The fitted stress sensitivity coefficient $\alpha$ is highest in the Northern region ($\alpha = 2.21$), intermediate in the Southern region ($\alpha = 1.44$), and lowest in the Central region ($\alpha = 0.94$). Corresponding results based on tidal normal stress (Figure S2 in Supporting Information) exhibit the same regional contrast for both $P$ and $\alpha$. These values fall within the range reported in previous studies of tidal triggering of tremor and earthquakes, where stress sensitivity is generally modest but detectable (e.g., \parencite{yabe2015tidal,hirose2025tidal}). However, we do not directly compare $\alpha$ values across studies, because $\alpha$ depends on the definition and amplitude of the imposed stress perturbation.

In addition to the resolved tidal shear and normal stresses, we also examine tidal sensitivity based on the stress invariants that are independent of fault orientation, including the mean stress $p$ and equivalent shear stress $q$ (see Text S1 and Figure S3 in Supporting Information). The same spatial ordering is preserved across all stress representations, with the Northern region consistently showing the strongest modulation and the Central region the weakest. This persistence indicates that the observed along-strike variation in tidal sensitivity is robust.

\subsection{Spatial variations of tremors, earthquakes, and tidal sensitivity}
\label{sec_spatial_variation}
\begin{figure}
    \centering
    \includegraphics[width=0.9\linewidth]{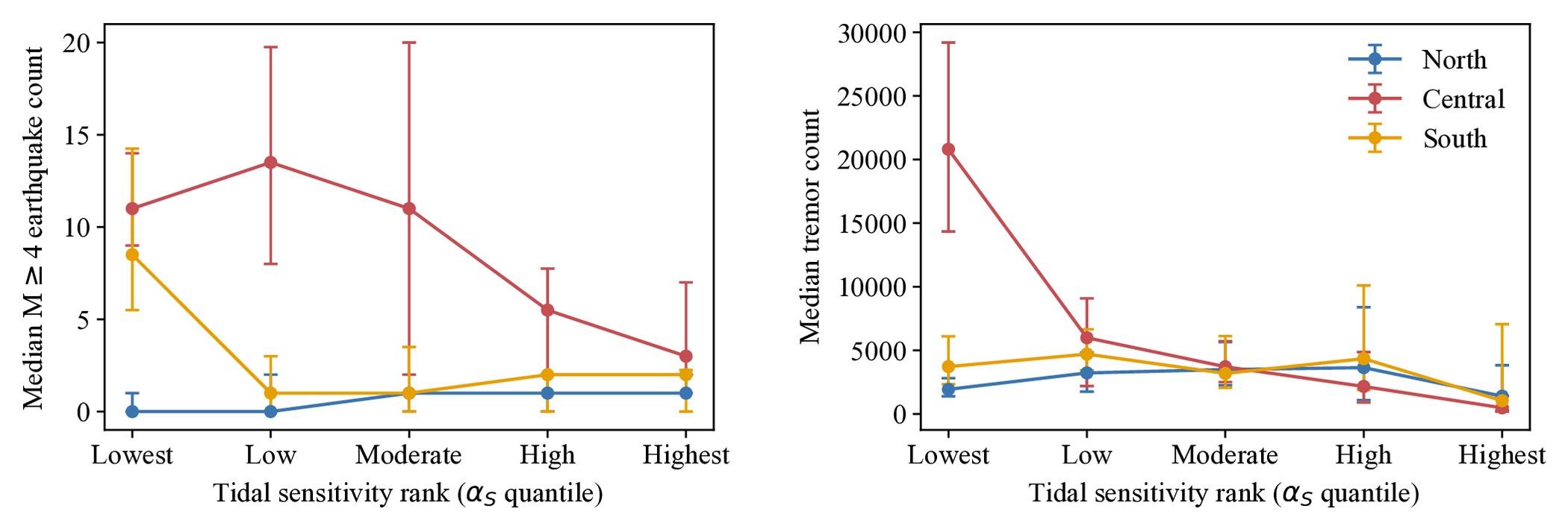}
    \caption{
    Quantitative relationships between tidal sensitivity, earthquake occurrence, and tremor occurrence.
    Tidal sensitivity $\alpha_S$ was ranked within each region and divided into five equal-sized bins from lowest to highest.
    Symbols denote the median earthquake or tremor count in each bin, and error bars denote the interquartile range across grid cells.
    Earthquake counts are calculated for nearby fast earthquakes with $M_j \geq 4$.
    }
    \label{fig:tidal_sensitivity_quantitative}
\end{figure}

The regional analysis above shows clear along-strike differences in tidal sensitivity, but it does not resolve whether these differences are expressed uniformly within each region or are controlled by more localized variations. We therefore perform a grid based analysis to examine the spatial relationship among tidal sensitivity, seismicity, and tremor occurrence(Figure~\ref{fig:tidal_sensitivity_quantitative}).
The study area is sampled on a regular latitude longitude grid with a spacing of $0.1^\circ$. At each grid point, $\alpha_S$ is estimated using tremors located within a 20~km radius around that point. We retain only grid points with at least 100 tremors and positive tidal sensitivity that exceeds its uncertainty ($\alpha_S > \delta\alpha_S$). For the retained grid points, we also count earthquakes with $M_j \geq 4$ and tremors within the same 20~km radius, allowing direct spatial comparison among tidal sensitivity, seismicity, and tremor occurrence.

\begin{figure}
    \centering
    \includegraphics[width=\linewidth]{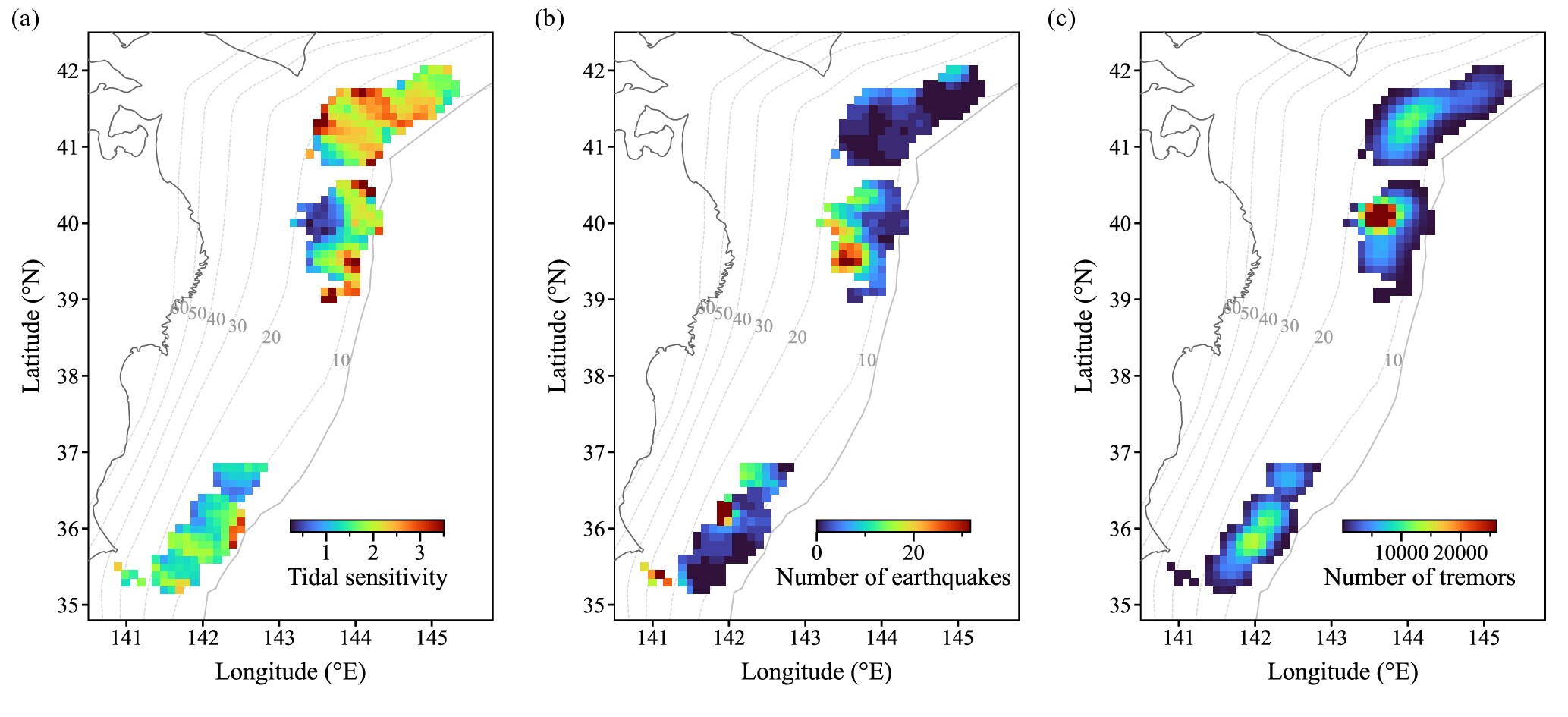}
    \caption{
    Spatial relationship among tidal sensitivity, seismicity, and tremor occurrence.
    (a) Spatial distribution of tidal sensitivity $\alpha_S$ from a grid based analysis. 
The map is sampled every $0.1^\circ$ in latitude and longitude, and the value at each grid point is calculated using tremors located within a 20-km radius around that point. Only points satisfying the selection criteria for the $\alpha_S$ analysis are shown. Results obtained using a 10-km radius are provided in the Supplementary Material.
    (b) Number of nearby earthquakes ($M_j \geq 4$) evaluated at the same grid points.
    (c) Number of tremors evaluated at the same grid points.
    }
    \label{fig:seismicity_tidal_sensitivity}
\end{figure}

The map of $\alpha_S$ (Figure~\ref{fig:seismicity_tidal_sensitivity}a) shows pronounced spatial variations. Weaker tidal sensitivity is mainly observed in the western part of the Central region and the northwestern portion of the Southern region. Similar spatial patterns are obtained for $\alpha$ based on tidal normal stress and for the phase-based modulation amplitude $P$ (Figure S4 in Supporting Information). In contrast, earthquakes with $M_j \geq 4$ is relatively more in the western part of central and the northwestern portion of Southern regions (Figure~\ref{fig:seismicity_tidal_sensitivity}b), broadly corresponding to areas of weaker tidal sensitivity. The corresponding distribution using a lower magnitude threshold ($M_j \geq 2$) for comparison (Figure S5 in Supporting Information). The spatial distribution of tremor occurrence also exhibits strong regional variability (Figure~\ref{fig:seismicity_tidal_sensitivity}c), with particularly more tremors in the western part of the central region. 

To further check these spatial relationships, we rank $\alpha_S$ within each region and compare earthquake and tremor occurrence across different tidal sensitivity levels (Figure~\ref{fig:tidal_sensitivity_quantitative}). The binned comparison further highlights regional differences (Figure~\ref{fig:tidal_sensitivity_quantitative}). For each region, grid points are divided into four quantile bins, from the weakest to strongest tidal sensitivity. Because the ranking is performed separately within each region, the bins represent relative tidal sensitivity levels. For each bin, we calculate the median number of $M_j \geq 4$ earthquakes and the median number of tremors across grid points; error bars denote the interquartile range.
The binned comparison highlights clear regional differences. In the Central region, grid points with lowest tidal sensitivity tend to have more moderate sized earthquakes. This tendency is less evident in the Southern region and is not observed in the Northern region, where earthquake counts remain low across all $\alpha_S$ bins. Tremor occurrence shows a similar pattern, with the clearest contrast between weak and strong tidal sensitivity bins in the Central region. These results suggest that the inverse relationship between tidal sensitivity and earthquake or tremor occurrence is mainly expressed in the Central region.

\section{Detailed characterization of regional tremor behaviors}

\subsection{Tremors associated with earthquakes}
\label{sec_eq_tr}
The spatial comparison in Section~\ref{sec_spatial_variation} shows that areas of weaker tidal sensitivity coincide with regions of more moderate sized earthquakes and dense tremor activity, especially in the Central region. This spatial correspondence motivates a closer examination of how tremor evolves around nearby earthquakes. We therefore compare tremor associated with earthquake in the Central and Southern regions. The Northern region is not included in the main comparison because seismicity is relatively weak, although isolated cases of tremor increase following nearby earthquakes are observed there as well (Figure S6 in Supporting Information).

\begin{figure}[!h]
    \centering
    \includegraphics[width=0.7\linewidth]{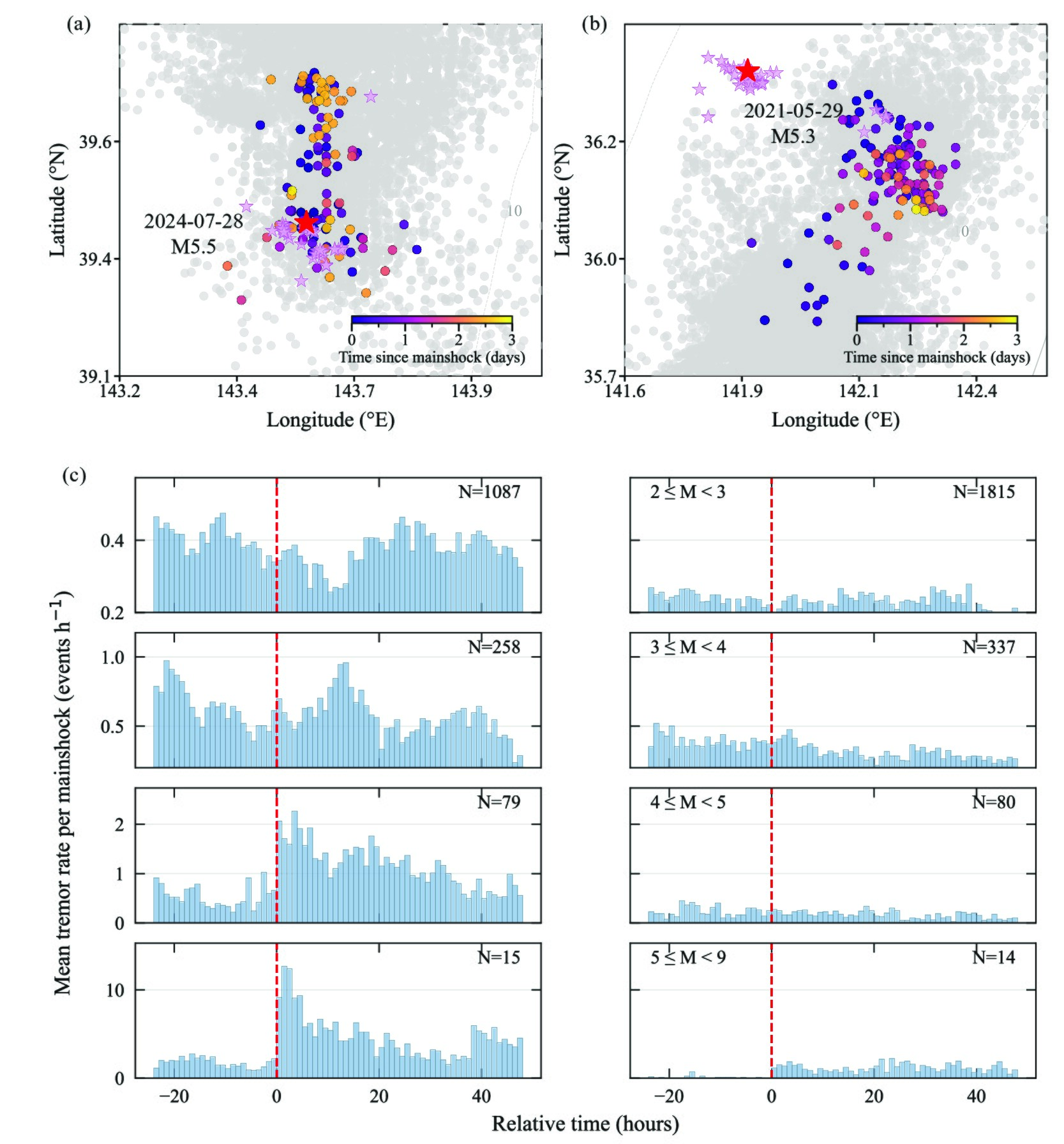}
    \caption{
    Representative examples of earthquake associated tremors and regional differences in tremor response to nearby earthquakes.
    (a,b) Representative cases in the central and Southern regions (corresponding to Figure~\ref{fig:spatiotemporal}e,f), showing tremor locations colored by time since the mainshock.
    Gray dots indicate background tremors before the earthquake.
    Red and pink stars denote the mainshock and aftershocks, respectively.
    (c) Average tremor rate as a function of time relative to nearby earthquakes in the central and Southern regions, grouped by earthquake magnitude.
    Tremor events within 60~km of each earthquake are included, and the tremor rate is averaged over all events in each magnitude range.
    The red dashed line marks the earthquake origin time.
    }
    \label{fig:eq_trigger_tr}
\end{figure}

We first compare representative examples from the central and Southern regions (Figure~\ref{fig:eq_trigger_tr}a,b). 
In the Central region, earthquakes tend to occur within the tremor region and are followed by spatially distributed tremor activation. In contrast, earthquakes in the Southern region more commonly occur near the edges of the tremor region, and the following tremor activity remains mainly confined within the tremor zone.

We further quantify the tremor rate to nearby earthquakes as a function of earthquake magnitude in the two regions separately (Figure~\ref{fig:eq_trigger_tr}c). For each earthquake, tremors within a distance of 60~km are included. Only earthquakes that are not preceded by a larger event within 24~h are considered.
The results reveal clear regional differences in the magnitude dependence of earthquake associated tremors. 
In the Central region, the average tremor rate increases following moderate sized earthquakes ($M_j \geq 4$), with a stronger response after larger events. 
In contrast, the Southern region shows a clear tremor increase mainly after larger earthquakes ($M_j \geq 5$), whereas earthquakes with $4 \leq M_j < 5$ do not produce a comparable response.

A consistency check using a shorter interaction distance of 20~km (Figure S7 in Supporting Information) shows similar patterns. 
Another test restricted to the period after September 2020 yields consistent results (Figure S8 in Supporting Information), indicating that these observations are not sensitive to potential changes in catalog completeness. 
This period corresponds to the incorporation of real time processing of waveform data from the S-net ocean bottom seismic network into the JMA catalog, which improves the detection capability and hypocenter accuracy of offshore earthquakes \parencite{nanjo2026non}.

\subsection{Persistent contrast between weak and strong tidal sensitivity areas in the Central region}

\begin{figure}
    \centering
    \includegraphics[width=0.9\linewidth]{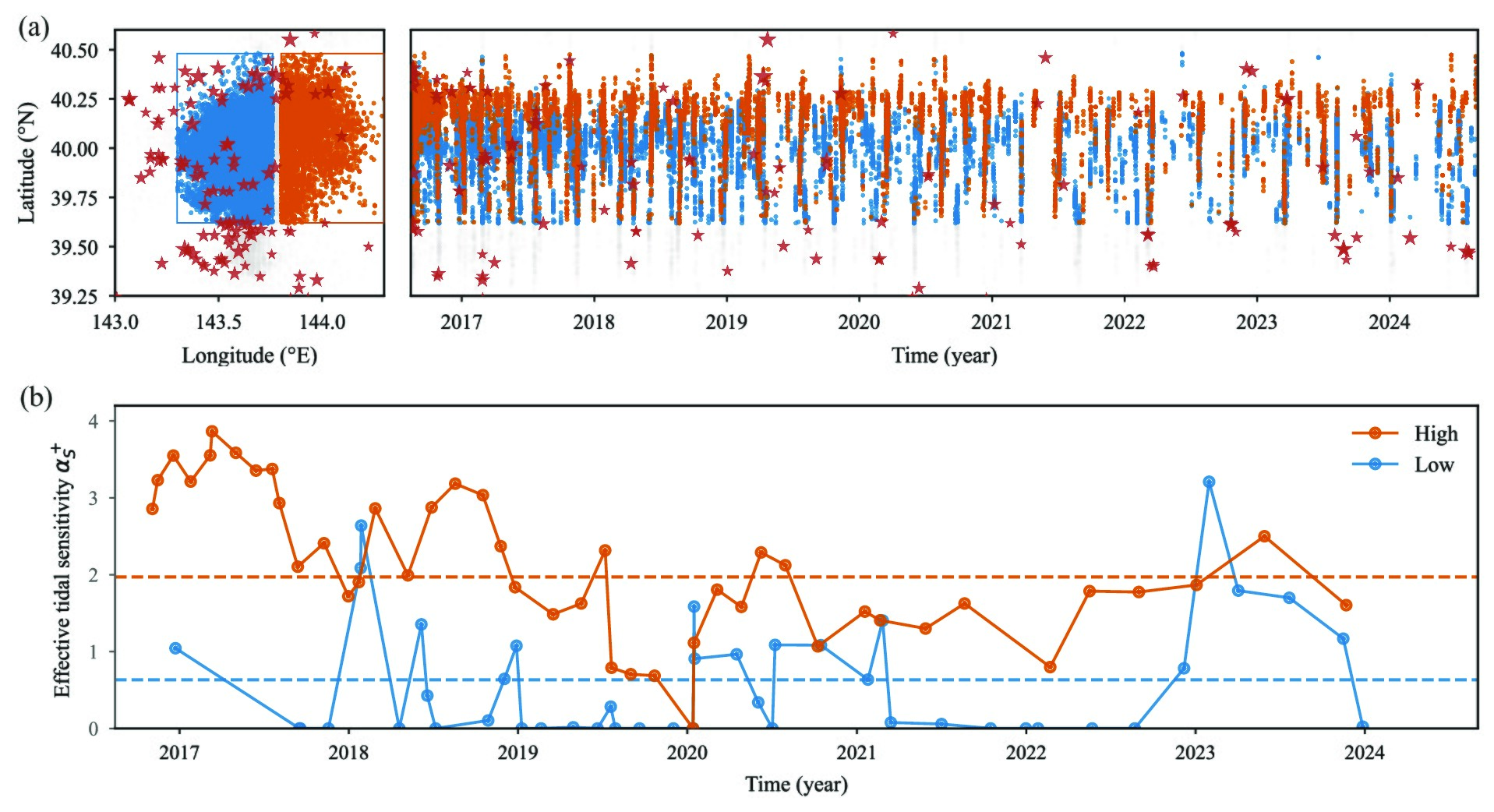}
\caption{
Spatiotemporal comparison of areas with weak and strong tidal sensitivity in the Central region.
(a) Map view of tremors in the western area with weaker $\alpha_S$ values (blue), the eastern area with stronger $\alpha_S$ values (green), and remaining tremors (gray). Red stars denote earthquakes with $M_j \geq 4$. Rectangles outline the two selected areas.
(b) Time latitude distribution of tremors and earthquakes in the Central region using the same color scheme as in (a). 
(c) Temporal evolution of $\alpha_S$ in the weak and strong tidal sensitivity areas estimated using fixed event moving windows. 
Dashed horizontal lines mark the duration weighted mean of $\alpha_S^{+}$ for each area.
}
    \label{fig:central_low_high}
\end{figure}

To examine whether this spatial contrast persists through time, we compare two representative areas from Figure~\ref{fig:seismicity_tidal_sensitivity} in the Central region: a western area with weaker tidal sensitivity and an eastern area with stronger tidal sensitivity (Figure~\ref{fig:central_low_high}).

The map and time latitude distributions (Figure~\ref{fig:central_low_high}a,b) show that tremors in the two areas remain spatially separated throughout the study period. The western area is characterized by denser tremor occurrence and more moderate sized earthquakes, whereas the eastern area shows sparser tremor activity and fewer moderate sized earthquakes.

We then examine the temporal evolution of $\alpha_S$ in the two areas using fixed event moving windows (Figure~\ref{fig:central_low_high}c). To account for the different tremor occurrence in the two areas, we use 2000 tremors per window with a step of 500 tremors for the western area, and 700 tremors per window with a step of 175 tremors for the eastern area. The eastern area maintains stronger tidal sensitivity throughout most of the study period, whereas the western area generally remains weaker and exhibits larger temporal fluctuations. Quantitatively, the duration weighted mean of $\alpha_S^{+}$ is stronger in the eastern area than in the western area (1.97 versus 0.63), while the weighted coefficient of variation is greater in the western area than in the eastern area (1.34 versus 0.40).

\subsection{Changes in tidal sensitivity during northern tremor migration}

\begin{figure}
    \centering
    \includegraphics[width=0.9\linewidth]{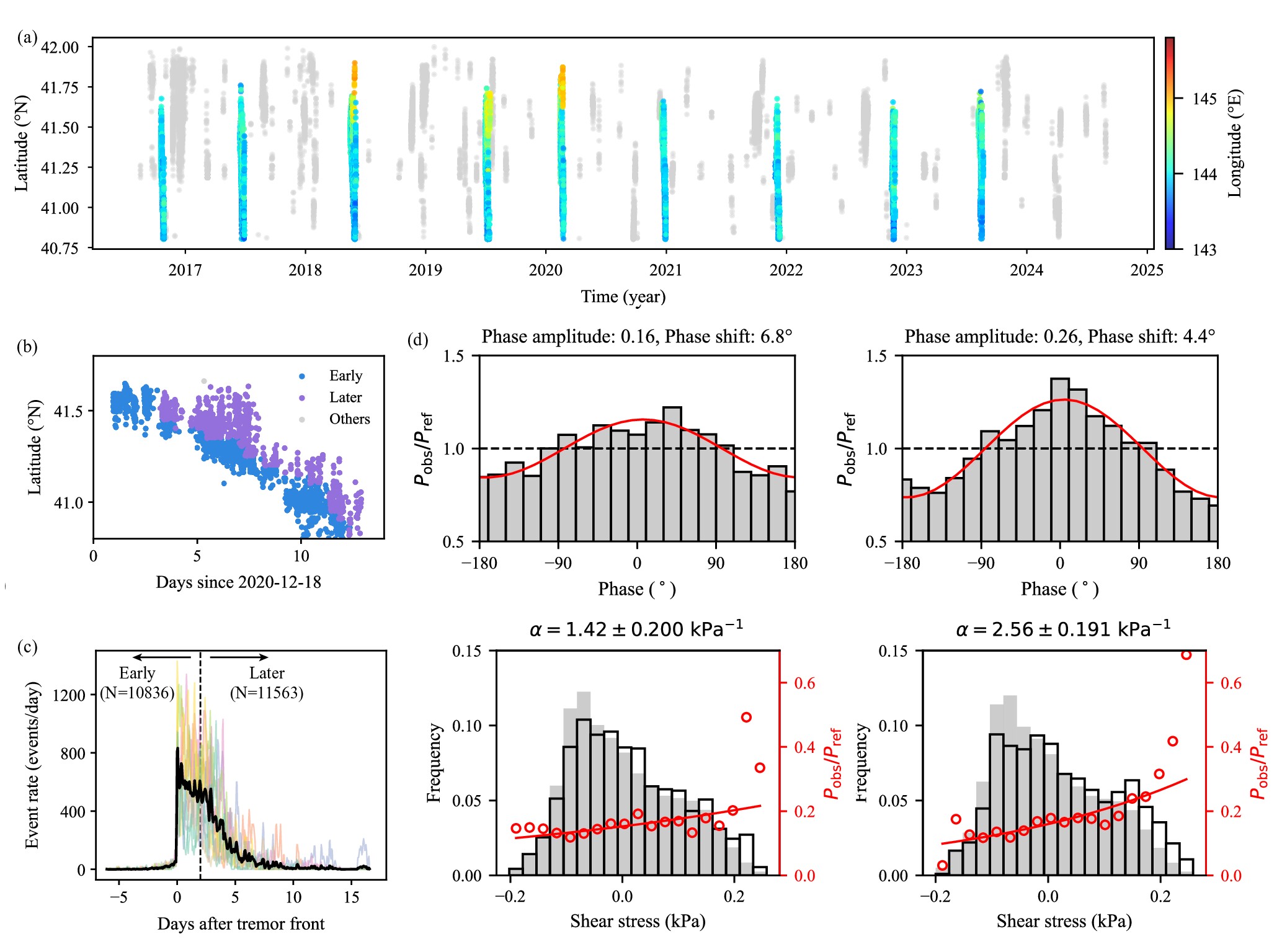}
    \caption{
Temporal evolution of tidal sensitivity during tremor migration in the Northern region. 
(a) Time-latitude distribution of tremors in the Northern region, highlighting nine extracted migrating clusters. Tremors within these clusters are colored by longitude, whereas all other tremors are shown in gray.
(b) Example of early- and later-stage classification for one representative cluster.
(c) Tremor rate aligned to the tremor-front time for the nine migrating clusters. Colored curves show the tremor-rate evolution for individual clusters, and the black curve shows the stacked tremor rate. The dashed line marks the boundary between the early and later stages, and the numbers of tremors assigned to each stage are indicated.
(d) Tidal sensitivity of the early and later stages. Upper panels show the phase dependence of tremor occurrence, and lower panels show the dependence on tidal shear stress amplitude.}
    \label{fig:ets}
\end{figure}

To investigate how tidal sensitivity evolves during tremor migration, we focus on migrating tremor clusters in the Northern region (as shown in Figure~\ref{fig:ets}a). Tremors in each migration are classified according to their occurrence time relative to the tremor front. The tremor front is defined in each $0.05^\circ$ latitude bin as the time when the tremor rate first exceeds 30 events per day. Then we set the boundary at 2 days after the tremor front: events occurring before this boundary are assigned to the early stage, whereas those occurring after it are assigned to the later stage. Tremors that cannot be assigned relative to an identified tremor front are excluded from the stage-based analysis. An example of this classification is shown in Figure~\ref{fig:ets}b, and the classification for all clusters is shown in Figure S9 in Supporting Information. The stacked temporal evolution of tremor migrations relative to their tremor front is summarized in Figure~\ref{fig:ets}c. Tidal sensitivity is then evaluated separately for the early and later stages. Figure~\ref{fig:ets}d shows that tremor occurrence in the later stage exhibits stronger tidal modulation than in the early stage. In the phase-based analysis, the later stage has a larger modulation amplitude ($P = 0.26$) than the early stage ($P = 0.16$). In the amplitude-based analysis, the stress sensitivity coefficient is also higher in the later stage ($\alpha = 2.56$) than in the early stage ($\alpha = 1.42$). Together, these results indicate that tidal sensitivity increases as tremor migration proceeds within the northern clusters, suggesting that the later stage of migration is more responsive to tidal perturbation. This stage-dependent increase in tidal sensitivity is consistent with previous observations from the Nankai and Cascadia subduction zones \parencite{thomas2013evidence,yabe2015tidal,houston2015low,peng2015high,katakami2017tidal}.

\section{Discussion}
%\subsection{Tectonic context of regional variations in tremor, earthquakes, and tidal sensitivity}

The three regions analyzed in this study show distinct patterns of tremor, earthquake occurrence, and tidal sensitivity. These regional differences are summarized in Table~\ref{tab:summary_regions}, together with independent tectonic and structural observations that may contribute to the regional differences. 

First, structural heterogeneity may provide a first-order control on the spatially variable tremor occurrence and tidal sensitivity. In the Central region, seismic imaging reveals strong variations in incoming plate faulting and sediment properties, including heterogeneous normal fault structures and variable sediment distribution \parencite{fujie2023nature,nakamura2023incoming}. Such structural complexity may influence local stress conditions and fluid distributions, consistent with the spatially variable tremor occurrence, coexistence of tremors and earthquakes, and weaker tidal sensitivity observed in this region. In the Southern region, the tremor distribution extends toward the northeastern limit of the Philippine Sea plate \parencite{uchida2009controls}, where the plate boundary geometry becomes more complex. Seismic imaging reveals pronounced structural heterogeneity, including bending-related normal faults, channel-like sedimentary structures, and nearby subducting seamounts \parencite{nakamura2023incoming,tsuru2002along,mochizuki2008weak}. These structural features may contribute to heterogeneous stress and fluid conditions, which may be related to the segmented tremor distribution and intermediate tidal sensitivity observed in this region.

Second, independently inferred aseismic slip processes differ among the three regions and may help explain their distinct tremor behavior. The Northern region ($\sim$40.8--42$^\circ$N) overlaps with the afterslip area of the 2003 Mw 8.0 Tokachi-oki earthquake and regions where potential SSEs have been reported \parencite{itoh2019interplate,okada2025investigation}. These independently inferred aseismic slip processes support the interpretation that the observed tremor migration may be driven by underlying slow slip processes. The Central region has also hosted transient aseismic slip, including the 1992 Sanriku-Oki ultra slow earthquake (Ms 6.9; $\sim$39.4$^\circ$N) \parencite{kawasaki19951992} and another one in early 2015 (39.2--40.2$^\circ$N; Mw $\sim$7.3) inferred from GNSS observations and repeating earthquake activity \parencite{honsho2019offshore,fujiwara2022spatiotemporal}. These events have been interpreted as earthquake-related slow slip phenomena with characteristics intermediate between typical SSEs and afterslip \parencite{nishikawa2023review}. In the Southern region (35--36.8$^\circ$N), multiple shallow SSE candidates have been reported \parencite{nishikawa2023review,nishimura2021slow}, and some tremor migration episodes are broadly consistent in space and time with independently inferred SSEs \parencite{sagae2025machine}, suggesting a possible link between tremors and SSEs. Thus, aseismic slip processes appear to be expressed differently among the three regions: coherent migration-related activity in the Northern region, earthquake-related transient slip in the Central region, and spatially distributed shallow SSEs in the Southern region.

Third, earthquake-related perturbations may contribute to the association between weaker tidal sensitivity, more moderate-sized earthquakes, and denser tremor activity, especially in the Central region. These perturbations may arise from processes operating over different timescales. At the longest timescale, differences in earthquake-cycle stage may contribute to along-strike contrasts in stress state. During the interseismic period, locked asperities on the megathrust can produce elastic loading in surrounding regions, leading to spatial variations in slip deficit and stress accumulation \parencite{herman2018accumulation}. The Central region is located updip of the 1968 Mw 8.3 Tokachi-oki earthquake rupture area, whereas the Northern region is closer to the more recent 2003 Tokachi-oki rupture area. Differences in accumulated interseismic loading may therefore contribute to their contrasting stress states. Following large earthquakes, postseismic perturbations may also affect tremor activity. The Central region lies near the northern margin of the 2011 Mw 9.0 Tohoku-Oki coseismic high-slip area and has experienced continued postseismic deformation since the 2011 earthquake \parencite{fukuda2021bayesian,honsho2019offshore,ozawa2012preceding,sagiya2022geodetic}. The gradual decrease in tremors from 2016 to 2024 may be related to the decay of postseismic perturbations, although separating this effect from longer-term earthquake-cycle processes remains difficult. At shorter timescales, tremor rate increases following some earthquakes, especially in the Central region. Similar earthquake-associated tremor activity has also been observed in other regions, where tremor activity increases following nearby moderate earthquakes \parencite{sagae2026detection,nadeau2009nonvolcanic}. Previous studies have also suggested possible interactions between nearby smaller earthquakes and tremor activity \parencite{farge2025big}. However, coseismic stress changes, aftershocks, and postseismic deformation may overlap in space and time, making it difficult to isolate the contribution of individual processes. Therefore, the association between weaker tidal sensitivity and moderate-sized earthquakes observed in this study may reflect cumulative and spatially distributed perturbations rather than a simple response to individual earthquake events.

\begin{table*}[t]
\centering
\caption{
Summary of regional differences in tremors, earthquakes, tidal sensitivity, and possible tectonic context along the Japan Trench.
}
\label{tab:summary_regions}
\renewcommand{\arraystretch}{1.35}
\setlength{\tabcolsep}{6pt}

\begin{tabularx}{\textwidth}{>{\raggedright\arraybackslash}p{3.8cm} 
                                >{\raggedright\arraybackslash}X 
                                >{\raggedright\arraybackslash}X 
                                >{\raggedright\arraybackslash}X}
\toprule
\textbf{Characteristic} & \textbf{North} & \textbf{Central} & \textbf{South} \\
\midrule

\multicolumn{4}{l}{\textit{Observations from this study}} \\
\midrule

Tremor migration
& Yes
& No 
& No \\

\addlinespace

Tidal sensitivity of tremor
& Highest (stronger during later migration)
& Lowest (nearly absent in the western part)
& Intermediate (weaker near western edges) \\

\addlinespace

EQ rate density (yr$^{-1}$ deg$^{-2}$)
& $\sim$40
& $\sim$170
& $\sim$260 \\

\addlinespace

Earthquake associated tremor pattern
& Rare increase after EQ (Fig.~S2)
& Clear response after $M_j \geq 4$ EQs within the tremor belt (Fig.~\ref{fig:eq_trigger_tr}a)
& Clear response mainly after $M_j \geq 5$ edge EQs (Fig.~\ref{fig:eq_trigger_tr}b) \\\\
\addlinespace

\midrule
\multicolumn{4}{l}{\textit{Independent tectonic context}} \\
\midrule

Structural characteristics
& Change in trench strike
& Strong structural heterogeneity \parencite{fujie2023nature,nakamura2023incoming}
& Complex structures including half-graben structure, fluid effects, and possible seamount influence \parencite{tsuru2002along,mochizuki2008weak,nakamura2023incoming} \\
\addlinespace

Aseismic deformation
& Afterslip and possible SSEs \parencite{okada2025investigation}
& Earthquake related SSEs \parencite{nishikawa2023review}
& Abundant SSEs \parencite{nishikawa2023review,nishimura2021slow} \\

\bottomrule
\end{tabularx}
\end{table*}

\section{Conclusions and implications}
We investigated the tidal sensitivity of shallow tectonic tremors along the northeastern Japan subduction zone. Tidal sensitivity shows clear along strike variations. It is strongest in the Northern region, weakest in the Central region, and intermediate in the Southern region. Two detailed observations further clarify these regional variations. First, in the Central region, areas with weaker tidal sensitivity are associated with more moderate sized earthquakes and denser tremor activity, whereas nearby areas with fewer earthquakes maintain stronger and more stable tidal sensitivity. This contrast persists through time, suggesting that tidal sensitivity reflects longer lived differences in local fault conditions or perturbation environments. Second, in the Northern region, tidal sensitivity becomes stronger during the later stages of tremor migration. Because tremor migration is commonly associated with slow slip processes, this increase suggests that the tidal response evolves as the tremor system changes during migration.

These results suggest that tidal sensitivity provides indirect information about the susceptibility of tremor generating regions to weak periodic stress perturbations. Stronger tidal sensitivity may indicate conditions in which small tidal stresses can more effectively modulate tremor occurrence, possibly because the fault is closer to failure or has lower effective strength. Weaker tidal sensitivity, however, does not necessarily imply that the fault is farther from failure. It may also indicate that tremor activity is influenced by larger, cumulative, or more heterogeneous perturbations, so that the tidal signal becomes less clearly expressed. Therefore, spatial and temporal variations in tidal sensitivity should be interpreted as a relative measure of how tremor activity responds to weak tidal perturbations in the presence of other ongoing stress perturbations.

Previous studies have established that tidal sensitivity reflects the susceptibility of tremor generating regions to weak periodic stress perturbations. Our results build on this view by suggesting that spatial and temporal variations in tidal sensitivity can help compare relative differences in perturbation environments among tremor regions within a single shallow tremor system. Stronger tidal sensitivity suggests that small tidal stresses are more clearly expressed in tremor occurrence. Weaker tidal sensitivity, however, does not necessarily imply a lower susceptibility to weak stress perturbations; it may instead reflect the influence of larger, cumulative, or more heterogeneous perturbations, although the specific processes cannot be uniquely identified from tidal sensitivity alone.

Overall, along strike variations in tidal sensitivity provide a useful comparative probe of tremor generating regions along the northeastern Japan subduction zone. These variations suggest that tremor activity is influenced differently across regions by slow slip processes, moderate sized earthquakes, and structural heterogeneity along the shallow plate interface.

\section*{Acknowledgments}
YZ, HA, HSB, SI, and AS gratefully acknowledge the support provided through the CNRS-University of Tokyo joint program SESAME. HSB gratefully acknowledge the European Research Council (ERC) for its partial support of this work through the PERSISMO grant (No. 865411). YZ thanks Weifan Lu from ISTerre for providing the tidal sensitivity analysis code, and Suguru Yabe from the Geological Survey of Japan, AIST, for providing the tidal stress calculation code. YZ also thanks Seiya Yano from the University of Tokyo, and Ankit Gupta and Suli Yao from ENS, for helpful discussions. Large language models (LLMs) were used only for language editing and grammatical corrections of the manuscript.

\section*{Data Availability Statement}
The tremor catalog used in this study is the Data Set S1 in the Supporting Information of \textcite{sagae2025machine}. The earthquake catalog used in this study is provided by the Japan Meteorological Agency (JMA). 
The unified JMA earthquake catalog up to the end of 2023 is available at 
\url{https://www.data.jma.go.jp/eqev/data/bulletin/hypo_e.html}. 
Earthquake data after 2023 are obtained from the JMA daily earthquake reports 
(\url{https://www.data.jma.go.jp/eqev/data/daily_map/index.html}).

\section*{Conflict of Interest}
The authors declare no conflicts of interest relevant to this study.

\clearpage
\setcounter{figure}{0}
\renewcommand{\thefigure}{S\arabic{figure}}
\setcounter{table}{0}
\renewcommand{\thetable}{S\arabic{table}}

\setcounter{section}{0}
\renewcommand{\thesection}{S\arabic{section}}
\renewcommand{\thesubsection}{S\arabic{section}.\arabic{subsection}}
 
\pagestyle{custom}

\begin{center}
\Large \textbf{Supporting Information for "Tidal sensitivity of tremors in a mixed fast and slow earthquake system in northeastern Japan"}\\[12pt]

\normalsize

Yishuo Zhou$^{1}$, 
Hideo Aochi$^{1,2}$, 
Alexandre Schubnel$^{1}$, 
Satoshi Ide$^{3}$,
Harsha S. Bhat$^{1}$

\begin{enumerate}
\small
\setlength\itemsep{-5pt}
\item {Laboratoire de G\'{e}ologie, Ecole Normale Sup\'{e}rieure, CNRS-UMR 8538, PSL Research University, Paris, France}
\item {Bureau de Recherches G\'{e}ologiques et Mini\'{e}res (BRGM), 45100 Orl\'{e}ans, France}
\item {Department of Earth and Planetary Science, The University of Tokyo, Tokyo, Japan}
\end{enumerate}
\end{center}

\section*{Contents of this file}
1. Text S1 \\
2. Figures S1-S9

\section*{ Text S1. Stress invariants independent of fault orientation}

To further evaluate whether the inferred along-strike variation in tidal sensitivity depends on the choice of stress measure, we additionally examine stress invariants that are independent of the assumed fault orientation. In particular, we consider the mean stress $p$, which characterizes the isotropic component of the stress perturbation, and the equivalent shear stress $q$, which characterizes the intensity of the deviatoric stress perturbation.

The mean stress is defined as
\begin{equation}
p = \frac{1}{3}\left(\sigma_{rr} + \sigma_{\theta\theta} + \sigma_{\phi\phi}\right),
\end{equation}
where $\sigma_{rr}$, $\sigma_{\theta\theta}$, and $\sigma_{\phi\phi}$ are the diagonal components of the tidal stress tensor in spherical coordinates.

The equivalent shear stress is defined from the second invariant of the deviatoric stress tensor,
\begin{equation}
q = \sqrt{3J_2},
\end{equation}
where
\begin{equation}
J_2 = \frac{1}{6}\left[
(\sigma_{rr}-\sigma_{\theta\theta})^2 +
(\sigma_{\theta\theta}-\sigma_{\phi\phi})^2 +
(\sigma_{\phi\phi}-\sigma_{rr})^2
\right]
+ \sigma_{r\theta}^2 + \sigma_{\theta\phi}^2 + \sigma_{\phi r}^2.
\end{equation}

Here, $p$ represents the isotropic part of the tidal stress perturbation, whereas $q$ provides a scalar measure of the shear intensity that does not depend on resolving stress onto a specified fault plane. These quantities therefore provide a useful robustness check on the regional pattern inferred from fault-oriented stress components such as the resolved shear and normal stresses.

Using the same phase-based analyses as in the main text, we calculate tidal sensitivity in the northern, central, and southern regions with respect to $p$ and $q$. The resulting phase modulation and stress sensitivity show the same first-order along-strike pattern as those obtained using resolved shear and normal stresses: the northern region exhibits the strongest tidal modulation, the central region the weakest, and the southern region intermediate behavior (Figure S6). This consistency indicates that the observed regional contrast in tidal sensitivity is robust and does not arise solely from the assumed focal mechanism or fault-orientation parameters used to resolve tidal stress onto the plate interface.

\begin{figure}[t]
    \centering
    \includegraphics[width=0.7\linewidth]{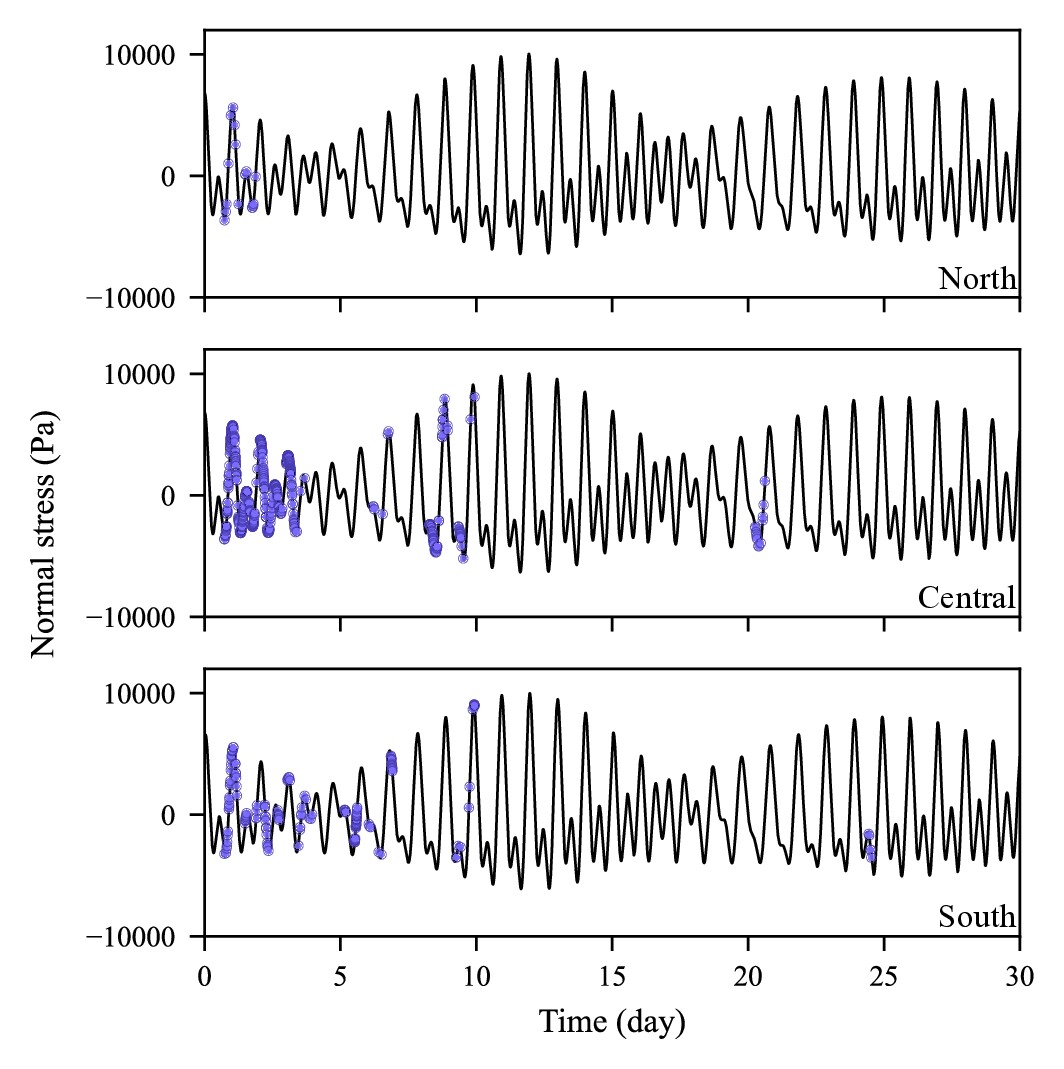}
    \caption{
Examples of the calculated tidal normal stress at the centers of the three study regions. 
The temporal variations are similar among regions, and the amplitudes are of order 10~kPa.
3}
    \label{fig:tidal_normal_stress}
\end{figure}

\begin{figure}[t]
    \centering
    \includegraphics[width=0.9\linewidth]{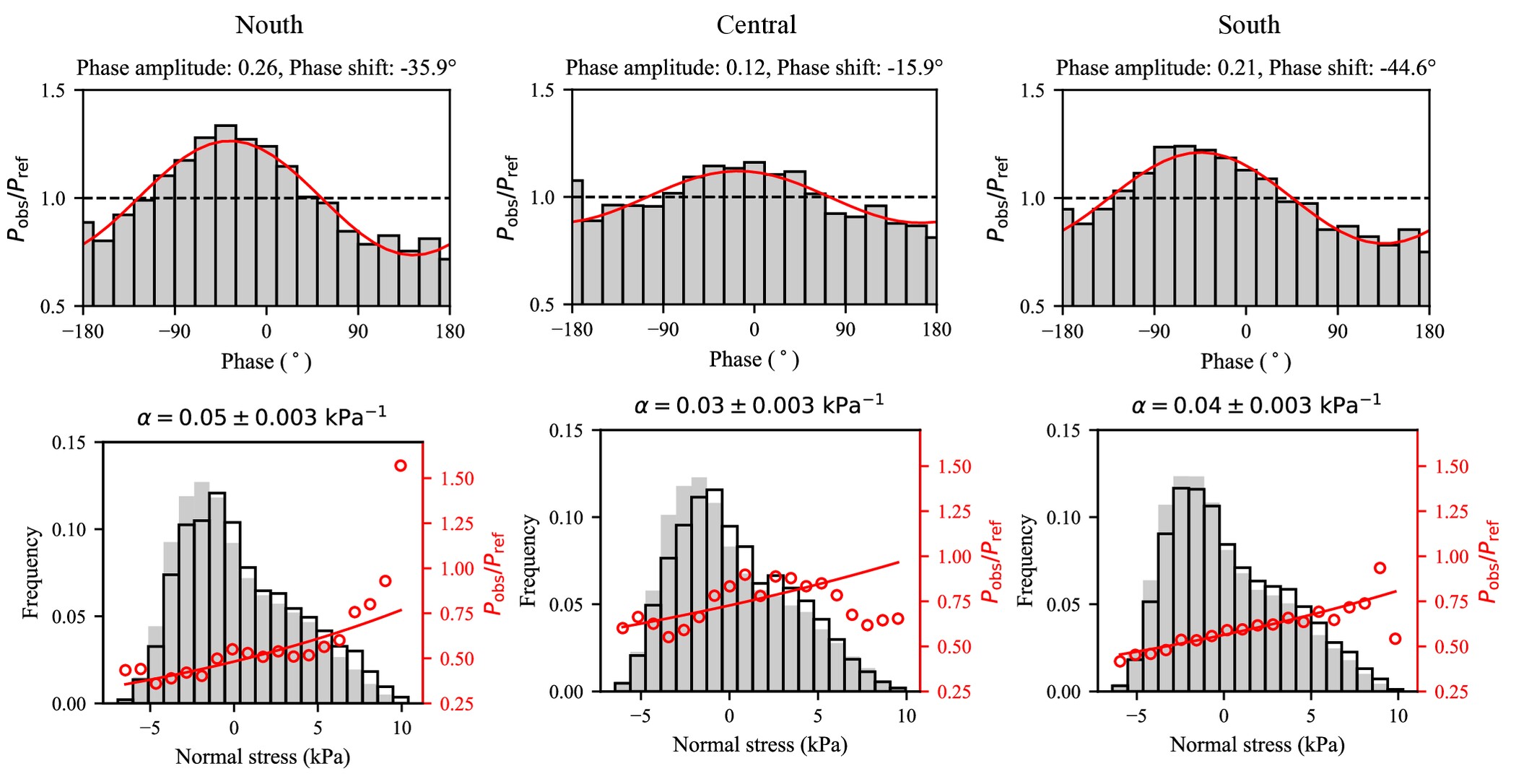}
    \caption{
Tidal sensitivity of tremor occurrence in the northern, central, and southern regions based on tidal normal stress. 
% The analysis is performed following the same procedure as in Figure 5, with tidal normal stress used instead of shear stress. 
Upper panels show the phase dependence of tremor occurrence, and lower panels show the dependence on stress amplitude. 
The results exhibit the same regional contrast as those based on tidal shear stress, with the northern region showing the strongest modulation, the central region the weakest, and the southern region intermediate behavior.
}
    \label{fig:S5}
\end{figure}

\begin{figure}[t]
    \centering
    \includegraphics[width=0.5\linewidth]{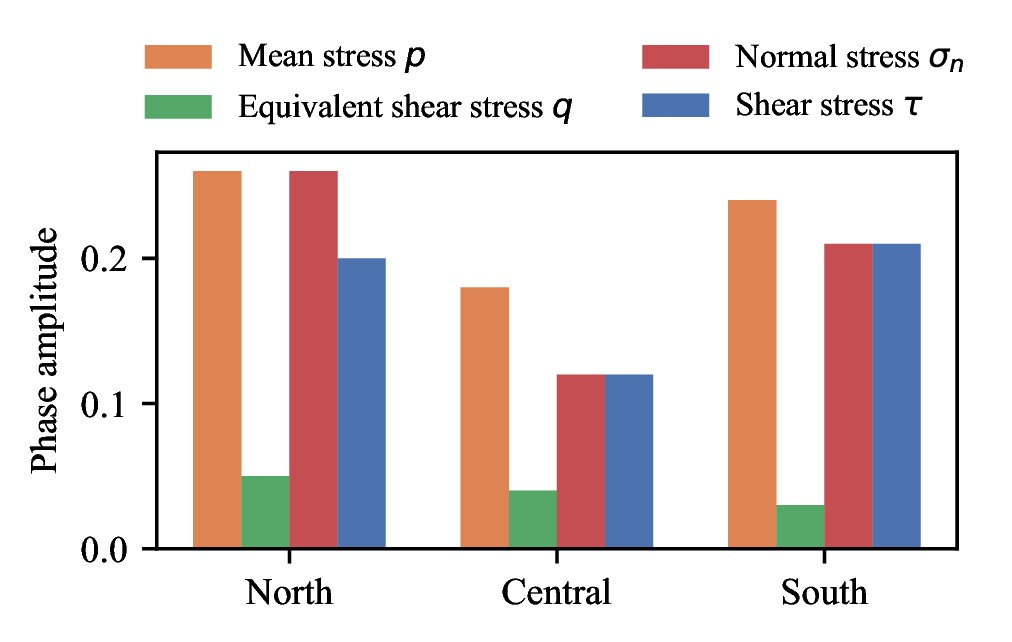}
    \caption{
Comparison of tidal modulation strength across different stress measures in the northern, central, and southern regions. 
Bars show the phase amplitude $P$ obtained from phase-based analysis using mean stress $p$, equivalent shear stress $q$, normal stress $\sigma_n$, and shear stress $\tau$. 
}
    \label{fig:S6}
\end{figure}

\begin{figure}[t]
    \centering
    \includegraphics[width=0.9\linewidth]{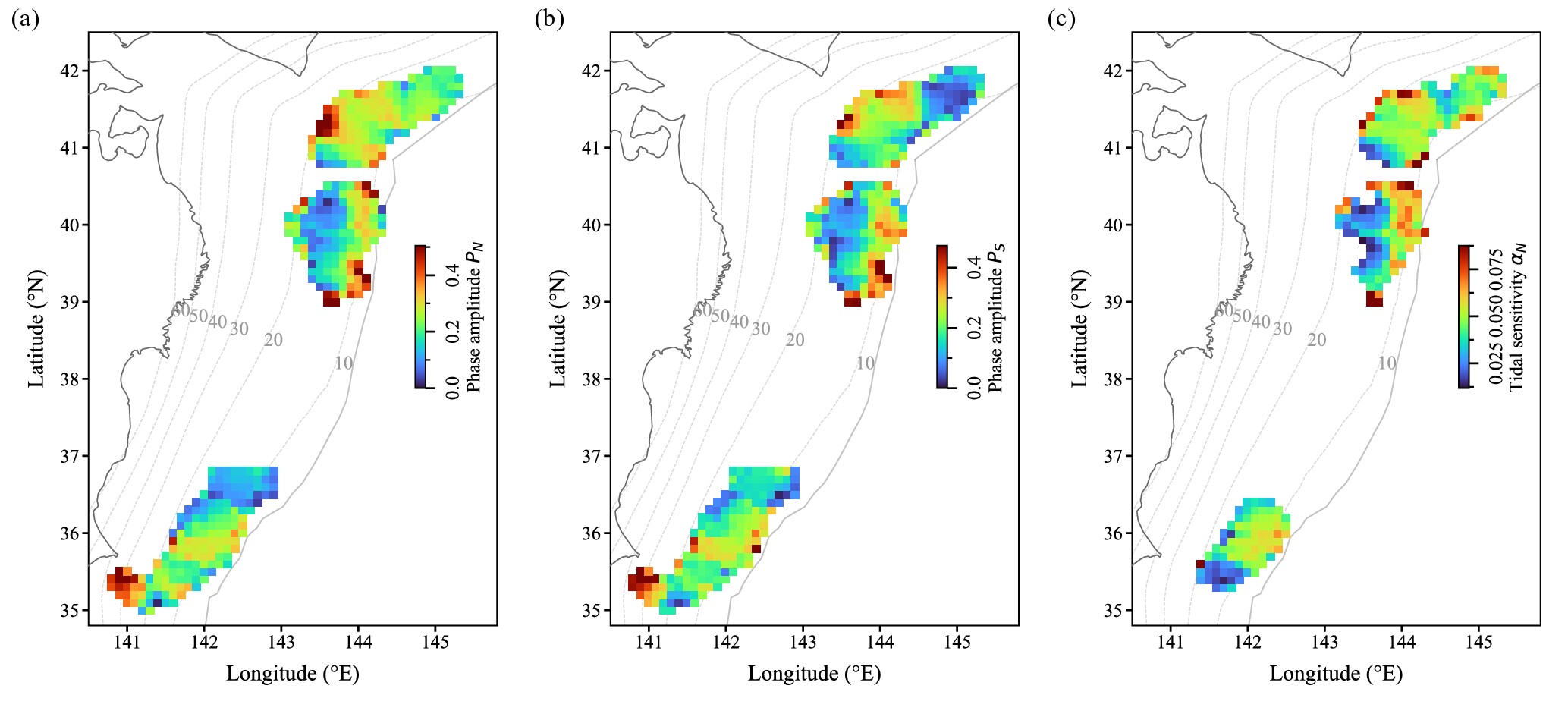}
    \caption{
Spatial distributions of additional tidal-sensitivity measures evaluated at analysis points spaced at $0.1^\circ$ intervals in latitude and longitude. 
At each point, the quantities are estimated using tremors located within a 20 km spherical radius, and only points satisfying the corresponding selection criteria are shown. 
(a) Phase amplitude based on tidal normal stress, $P_N$. 
(b) Phase amplitude based on tidal shear stress, $P_S$. 
(c) Stress sensitivity coefficient based on tidal normal stress, $\alpha_N$. 
}
    \label{fig:S8}
\end{figure}

\begin{figure}[t]
    \centering
    \begin{subfigure}{0.9\linewidth}
        \centering
        \includegraphics[height=0.43\textheight]{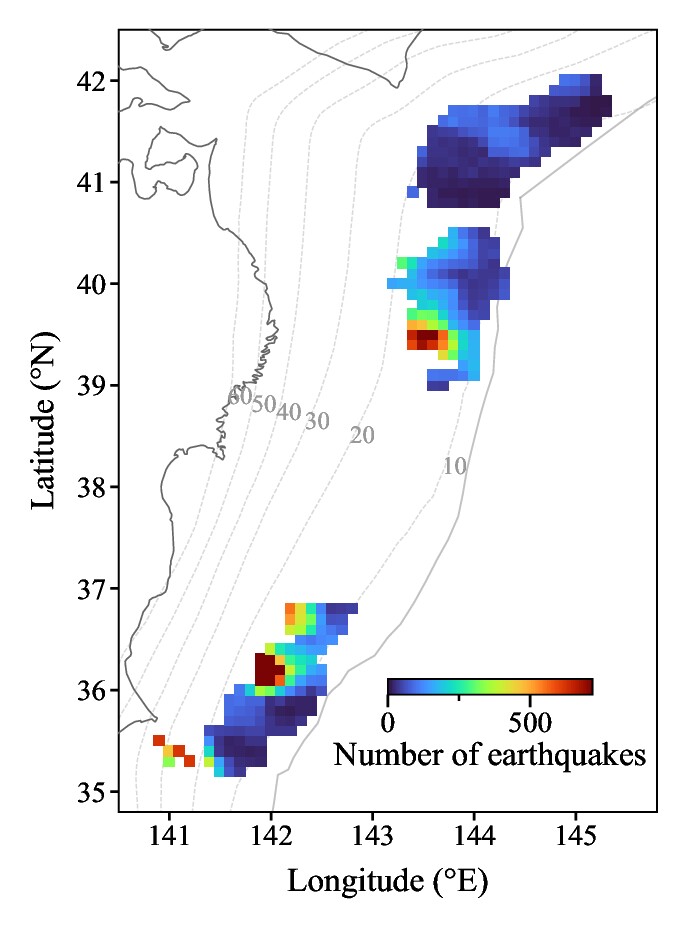}
        \caption{}
    \end{subfigure}

    \vspace{0.2cm}

    \begin{subfigure}{0.6\linewidth}
        \centering
        \includegraphics[height=0.33\textheight]{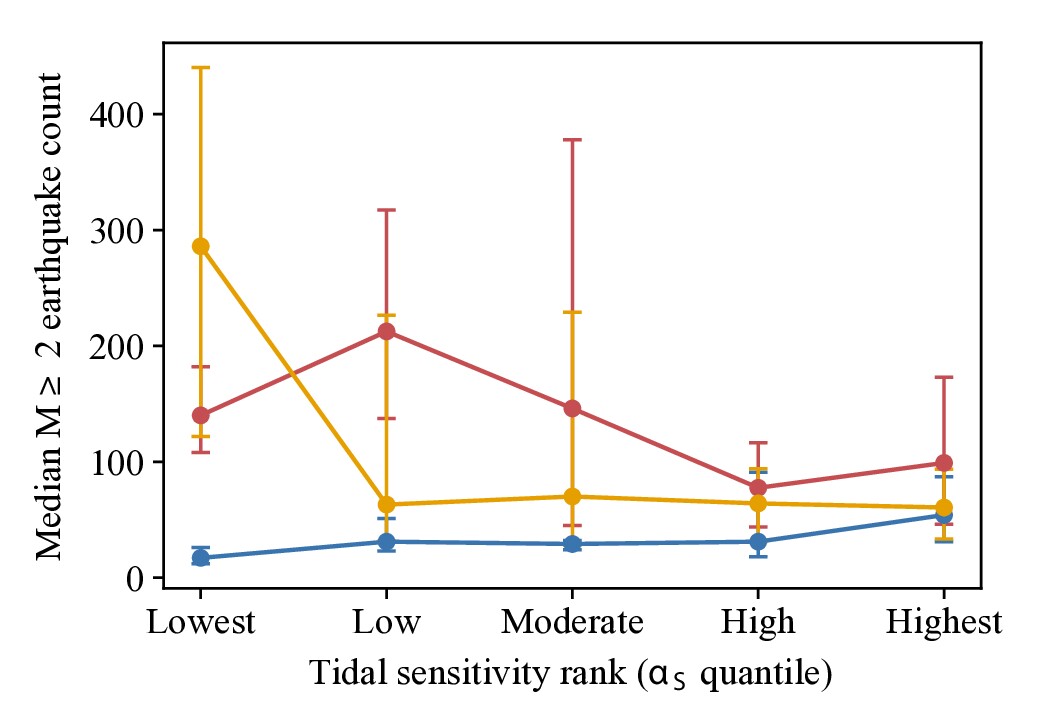}
        \caption{}
    \end{subfigure}

    \caption{
(a) Spatial distribution of earthquakes with magnitudes $M_j \geq 2$ evaluated on the same analysis grid as in Figure~5b.
(b) Relationship between tidal sensitivity $\alpha_S$ and earthquake activity based on the $M_j \geq 2$ catalog. Grid cells are grouped according to the quantiles of $\alpha_S$, and the corresponding earthquake counts are summarized for each group.
}
    \label{fig:S9}
\end{figure}

\begin{figure}[t]
    \centering
    \includegraphics[width=0.5\linewidth]{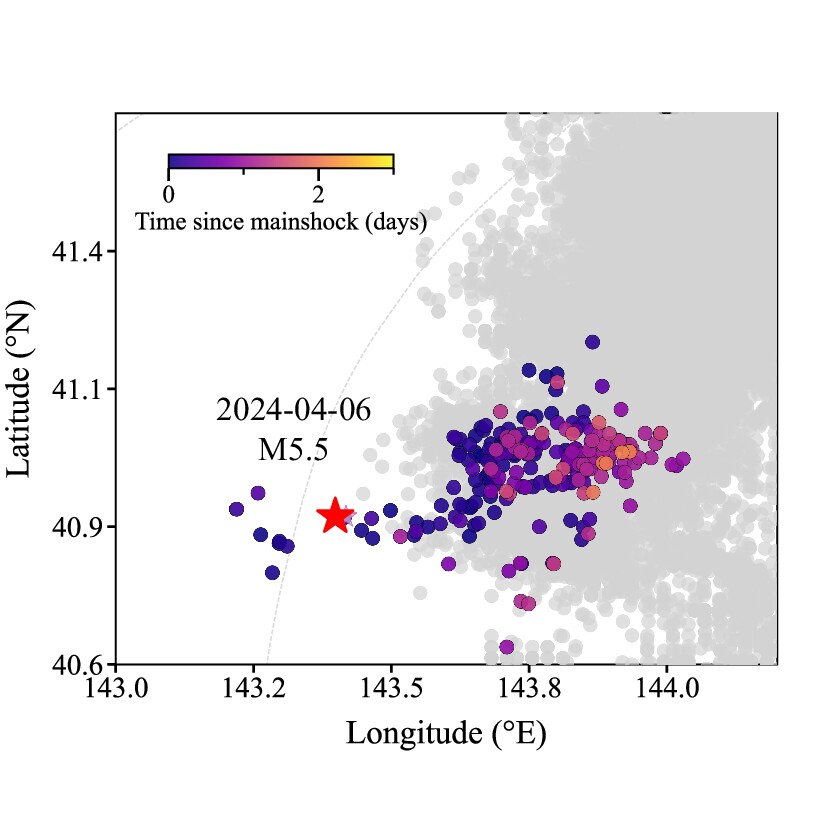}
    \caption{
Representative example of earthquake-associated tremor activation in the northern region. 
Gray symbols indicate tremor activity prior to the earthquake, and colored symbols indicate tremor activity after the earthquake, with color representing time since the mainshock. 
Although such cases can be identified, they are relatively rare and do not dominate the overall tremor pattern in the northern region.
}
    \label{fig:S2}
\end{figure}

\begin{figure}[t]
    \centering
    \includegraphics[width=0.9\linewidth]{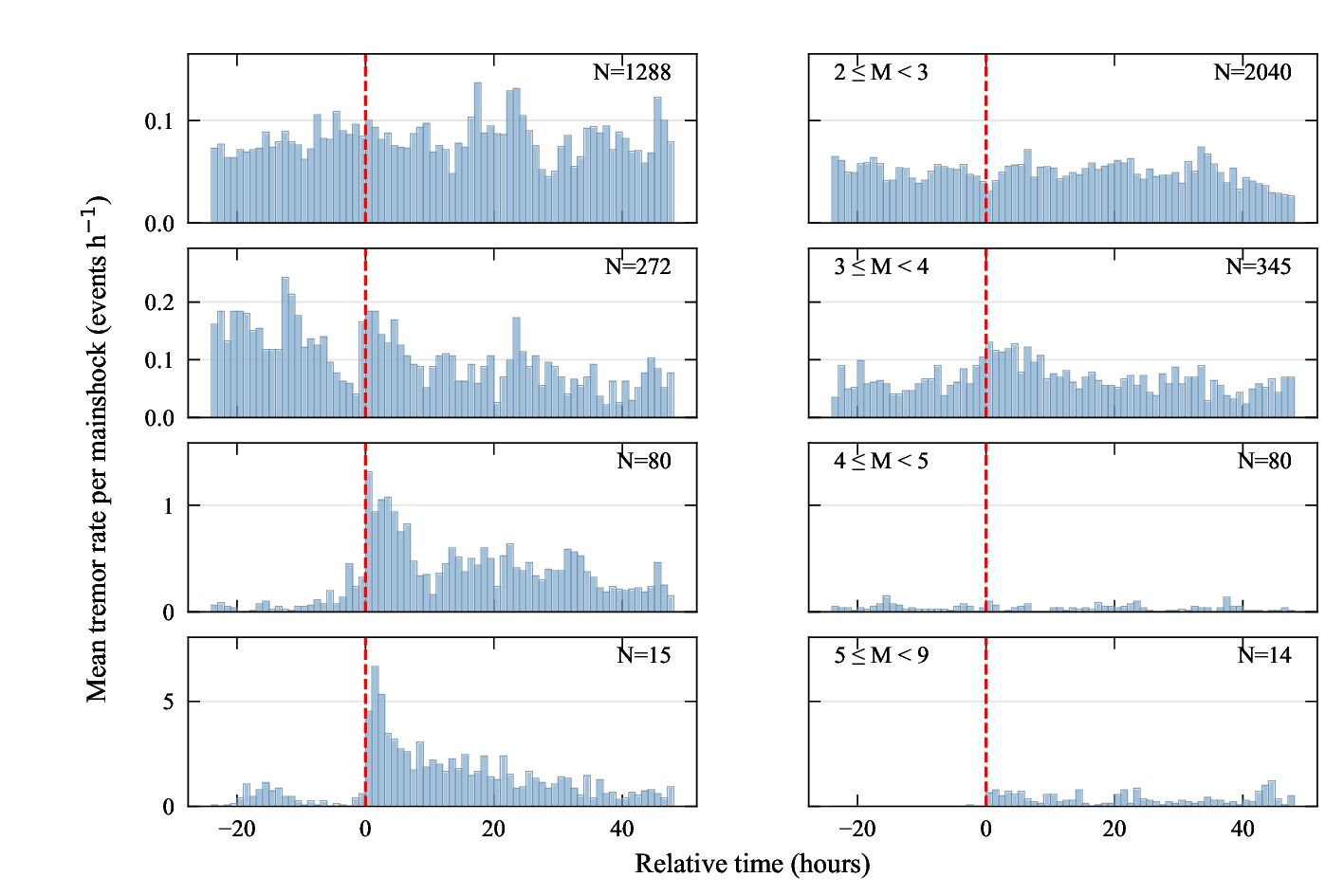}
    \caption{
Tremor response to earthquakes using a shorter interaction distance of 20~km. 
The average tremor rate is shown as a function of time relative to earthquake origin time, grouped by magnitude ranges. 
The overall pattern is consistent with that obtained using a 60~km interaction distance, with a clear increase in tremor rate following earthquakes with magnitudes $M_j \geq 4$.
}
    \label{fig:S3}
\end{figure}

\begin{figure}[t]
    \centering
    \includegraphics[width=0.9\linewidth]{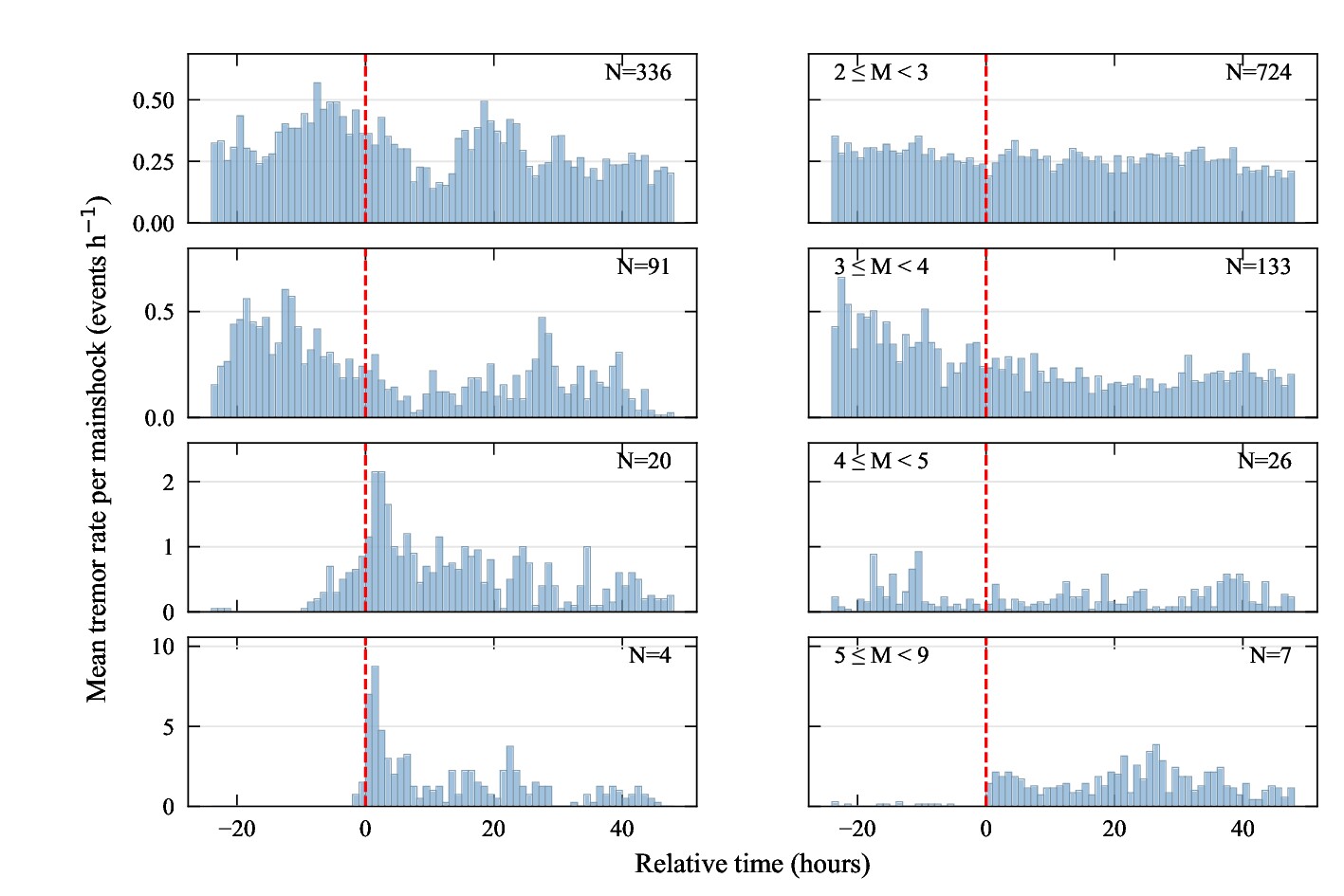}
    \caption{
Consistency check of tremor response restricted to the period after September 2020. 
Tremor rates are shown for interaction distances of 60~km. 
The magnitude-dependent pattern remains consistent with that obtained for the full study period, indicating that the results are not sensitive to potential variations in earthquake catalog completeness.
}
    \label{fig:S4}
\end{figure}

\begin{figure}[t!]
    \centering
    \includegraphics[width=0.9\linewidth]{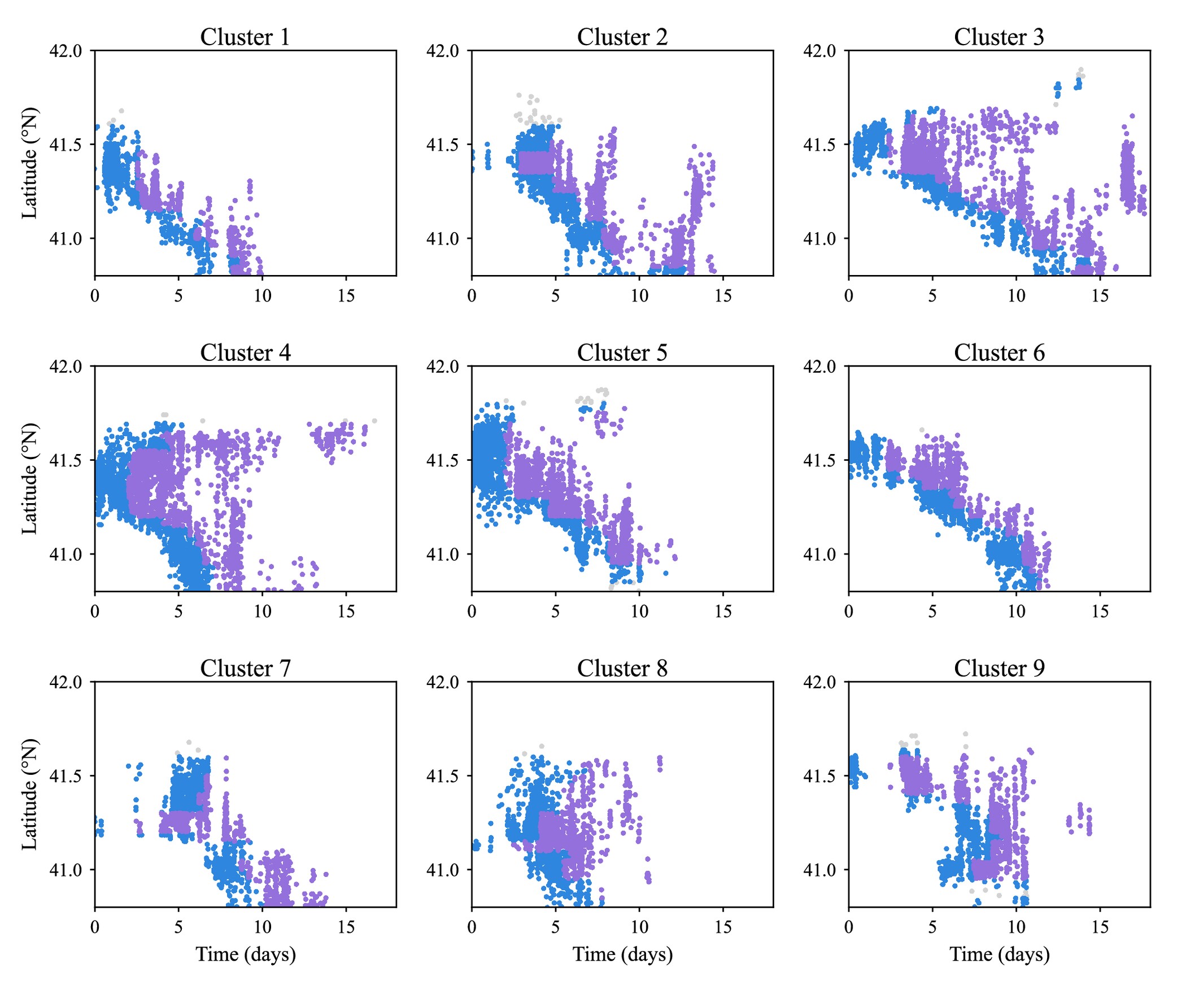}
    \caption{
Time–latitude distribution of tremor activity for the nine identified migrating clusters in the northern region. 
Each panel corresponds to one cluster. Tremor events are colored according to their classification into early and later stages relative to the tremor front, following the definition described in the main text. 
This figure illustrates the temporal structure of individual clusters and the basis for the stage-based classification used in the analysis of tidal sensitivity.
}
    \label{fig:S7}
\end{figure}

\clearpage
\printbibliography

\end{document}